\documentclass[journal]{mock-aiaa}
\usepackage[utf8]{inputenc}
\usepackage{graphicx,amsmath,longtable,tabularx,xcolor}
\usepackage{supertabular,booktabs}

\title{Meteor shower forecasting in near-Earth space}

\author{Althea V.~Moorhead\footnote{Aerospace Technologist, Planetary Studies; NASA Meteoroid Environment Office, MSFC EV44}} 
\affil{NASA Meteoroid Environment Office, Marshall Space Flight Center EV44, Huntsville, Alabama, 35812}
\author{Auriane Egal\footnote{Postdoctoral Fellow; Department of Physics and Astronomy, The University of Western Ontario}}
\affil{Department of Physics and Astronomy, The University of Western Ontario, London, Ontario N6A 3K7, Canada}
\author{Peter G. Brown\footnote{Professor; Department of Physics and Astronomy, The University of Western Ontario.}}
\affil{Department of Physics and Astronomy, The University of Western Ontario, London, Ontario N6A 3K7, Canada}
\affil{Centre for Planetary Science and Exploration, The University of Western Ontario, London, Ontario N6A 5B8, Canada}
\author{Danielle E.~Moser\footnote{Meteoroid Physicist; Jacobs, Jacobs Space Exploration Group, NASA Meteoroid Environment Office, MSFC EV44}}
\affil{Jacobs, Jacobs Space Exploration Group, Huntsville, Alabama, 35812}
\affil{NASA Meteoroid Environment Office, Marshall Space Flight Center EV44, Huntsville, Alabama, 35812}
\author{William J.~Cooke\footnote{Lead; NASA Meteoroid Environment Office, MSFC EV44}}
\affil{NASA Meteoroid Environment Office, Marshall Space Flight Center EV44, Huntsville, Alabama, 35812}

\begin{document}

\twocolumn[
  \begin{@twocolumnfalse}
    \maketitle

\begin{abstract}
NASA's Meteoroid Environment Office (MEO) produces an annual meteor shower forecast in order to help spacecraft operators assess the risk posed by meteoroid streams. Previously, this forecast focused on the International Space Station and therefore reported meteoroid fluxes and enhancement factors at an orbital altitude of 400 km. This paper presents an updated forecast algorithm that has an improved calculation of the flux enhancement produced by showers and can calculate fluxes at any selected Earth or lunar orbital altitude. Finally, we discuss and generate forecasted fluxes for the 2018 Draconid meteor shower, which is expected to produce meteoroid flux enhancements near the Sun-Earth L1 and L2 Lagrange points but not at Earth.
\end{abstract}

\vspace{0.333in}

  \end{@twocolumnfalse}
]

\section*{Nomenclature}
\tablefirsthead{} \tablehead{} \tabletail{} \tablelasttail{}
\begin{supertabular}{@{}l @{\quad=\quad} l@{}}
    $B_p$ & shower growth exponent, deg$^{-1}$ \\
    $B_m$ & shower decay exponent, deg$^{-1}$ \\
    BH & Brinell hardness \\
    $c$ & speed of sound, km~s$^{-1}$ \\
    $d$ & crater diameter, cm \\
    $f$ & meteoroid flux \\
    $G$ & gravitational constant \\
    $H$ & atmospheric thickness, km \\
    $h$ & altitude, km \\
    KE & kinetic energy, J \\
    $k_\textrm{avg}$ & perception factor \\
    $L$ & angular momentum \\
    $M$ & mass of a large body, kg \\
    $m$ & meteoroid mass \\
    $N$ & number of particles \\
    $p$ & crater depth in cm \\
    $R$ & average radius of a large body \\
    $r$ & population index or heliocentric distance \\
    $s$ & mass index \\
    $t$ & time \\
    $v$ & meteoroid velocity, km~s$^{-1}$ \\
    $x, y$ & position coordinates within the ecliptic plane \\ 
    ZHR & zenithal hourly rate, hr$^{-1}$ \\
    ZHR$_0$ & peak zenithal hourly rate \\
    $\alpha$ & shower flux fraction \\
    $\beta$ & flux enhancement factor \\
    $\gamma$ & damage rate multiplier \\
    $\delta$ & attitude disturbance rate multiplier \\
    $\zeta$ & attitude disturbance enhancement factor \\
    $\eta$ & gravitational focusing or shielding factor \\
    $\theta$ & impact angle, radians \\
    $\lambda_0$ & peak solar longitude \\
    $\lambda_\odot$ & solar longitude, deg \\
    $\xi$ & distance from center of spacecraft \\
    $\Pi$ & Heaviside pi function \\
    $\rho$ & density, kg~m$^{-3}$ \\
    $\phi$ & azimuthal angle, radians \\
    $\psi$ & planetary shielding angle, radians \\
    $\Omega$ & solid angle, steradians \\ 
\multicolumn{2}{@{}l}{}\\
\multicolumn{2}{@{}l}{\bf Subscripts}\\
    6.5 & for meteors brighter than magnitude 6.5 \\
    $b$ & of massive body $b$\\
    G & corresponding to \cite{1985Icar...62..244G} \\
    $g$ & gravitational focusing \\
    Gaia & of the Gaia spacecraft \\
    $h$ & at altitude $h$ \\
    $i$ & corresponding to individual shower $i$ \\
    ip & interplanetary \\
    lim & limiting \\
    mg & for meteoroids larger than 1~mg \\
    $\oplus$ & of the Earth \\
    $s$ & planetary shielding \\
    SOHO & of the SOHO spacecraft \\
    $t$ & target material \\
    TOA & at the top of the Earth's atmosphere \\
    TOB & at the top of a body's surface or atmosphere 
\end{supertabular}%

\section{Introduction}

\lettrine{M}{eteoroid} impacts pose a constant risk to spacecraft; their high speeds (12-72 km s$^{-1}$) compared to orbital debris means that a tiny particle can impart a great deal of kinetic energy, momentum, or damage.  The vast majority of the meteoroid flux is associated with the background component of the meteoroid environment: the so-called ``sporadic'' meteoroids. As a result, meteoroid environment models such as NASA's Meteoroid Engineering Model (MEM) \cite{2004EM&P...95..123M,Moorhead:2015wl} and ESA's Interplanetary Meteoroid Environment Model (IMEM) \cite{2005AdSpR..35.1282D} focus on the sporadic complex. Yet meteor showers can produce short-term enhancements of the meteoroid flux and associated risk that are not captured by models such as MEM and IMEM.  

Meteor showers can match or even exceed the sporadic flux for a short period; the Geminid meteor shower, for example, can double the meteoroid flux at the time of peak activity. The Leonid and Draconid meteor showers do not pose much of an impact risk in a typical year, but sometimes produce outbursts in which the level of activity can be tens to thousands of times higher than normal \cite{2014mesh.book.....K}.  Both showers produce ``storms'' -- i.e., meteor shower outbursts with a visual rate exceeding 1000 per hour \cite{2006ChAna..30...61M}. The phenomenal Leonid storm of 1833 had estimated hourly rates of 10,000 to 60,000 per hour \cite{2014mesh.book.....K}. The Draconid meteor shower is less well-known but also highly variable and has produced storms with rates of 10,000 per hour. In 2018, the Draconids were expected to produce high activity at the Sun-Earth L1 and L2 Lagrange points \cite{EgalLetter}.

Meteor showers are generally short-lived; durations can range from a few hours to a few weeks and, within those periods, most of the activity occurs near a ``peak'' time. This is particularly true for the Draconids, which tend to maintain at least half their maximum activity for only a few hours. As a result, spacecraft operators may choose to mitigate the risk operationally by phasing orbits, delaying launches, reorienting spacecraft, or powering down components. These mitigation techniques require accurate predictions of the timing and duration of meteor showers in order to be effective. NASA's Meteoroid Environment Office (MEO) therefore issues both annual and custom meteor shower forecasts that predict the timing and activity of meteor showers for particle sizes that are potentially threatening to spacecraft.

Our forecasts do not include the full working list of over 1000 meteor showers included in the International Astronomical Union (IAU) Meteor Data Center.\footnote{https://www.ta3.sk/IAUC22DB/MDC2007/index.php}
Only showers that are capable of producing an appreciable flux are incorporated; a typical forecast year is based on a list of around 30 meteor showers \cite{Moorhead2017}. The list of relevant showers must be reconsidered each year either as new information becomes available for poorly measured showers or as new predictions are made for variable showers. The MEO runs simulations of variable meteor showers each year in order to predict their activity \cite{2004EM&P...95..141M,2008EM&P..102..285M}; these predictions are fed into a forecasting algorithm that converts predicted visual rates to fluxes at a given spacecraft altitude \cite{Moorhead2017}.

The MEO forecasting code was originally designed to generate predictions for the Space Shuttle and International Space Station and thus computed meteoroid fluxes at an altitude of 400 km above the Earth's surface and for limiting particle kinetic energies that are relevant to a manned vehicle. In this paper, we present results from a significantly expanded version of the code that can generate meteoroid fluxes at any Earth-orbiting altitude. Furthermore, the code can compute fluxes on the Moon's surface or in lunar orbit, and can handle special locations such as the L1 and L2 Lagrange points. Our forecasts are also no longer restricted to calculating kinetic-energy-limited fluxes; we have modified the code to generate size- and mass-limited fluxes as well. Section \ref{sec:method} describes our improved forecasting algorithm.

Our forecasts do not simply report shower fluxes, but attempt to put these fluxes into context by comparing them with the sporadic or background meteoroid flux. We do so by providing ``enhancement factors'' that describe the factor by which the typical background flux is increased due to meteor showers. These enhancement factors may be combined with existing sporadic meteoroid risk assessments to estimate the increase in risk. We typically assume a ``worst-case scenario'' in which a spacecraft surface directly faces the shower radiant and does not benefit from any planetary shielding. However, depending on the response of the spacecraft material to meteoroid impacts, the risk may be enhanced to a greater degree than is described by our simple enhancement factors. In Section \ref{sec:xtrafac}, we compute the additional enhancement factors that would need to be applied for a sample ballistic limit equation (BLE) and for momentum disturbances.

Section \ref{sec:dra2018} reviews recent results from models of the 2018 Draconid meteor shower, which likely exhibited storm-level activity at both the L1 and L2 Sun-Earth Lagrange points. We derive a meteor shower profile from the results of Egal et al.~\cite{EgalLetter} and use our forecasting algorithm to calculate the equivalent fluxes. We present a suite of flux predictions for spacecraft orbiting near L1 and L2.

Note that in this paper, ``meteor'' and ``meteoroid'' carry different meanings. We use the term ``meteoroid'' to refer to a small natural particle moving through space, and the term ``meteor'' to refer to the light and ionization produced when a meteoroid enters the Earth's atmosphere. When we discuss fluxes, we are generally referring to meteoroids, and when we refer to hourly rates (especially zenithal hourly rates, or ZHRs), we are referring to meteors as seen by a ground-based observer. Although the terms ``micrometeoroid'' and ``micrometeorite'' are frequently used in the spacecraft engineering community, we opt to use the term ``meteoroid'' as micron- or microgram-sized particles are too small to pose a hazard to spacecraft.

\section{Forecast algorithm}
\label{sec:method}

This section describes our method for calculating meteor shower fluxes. We derive top-of-atmosphere fluxes as a function of time from a combination of meteor shower observations and numerical simulations. We then perform a series of calculations to extrapolate these fluxes to the desired orbital altitude and limiting quantity. Fluxes are also integrated over a time period of interest to obtain particle fluences. All fluxes and fluences are compared with the sporadic meteoroid flux and put in terms of enhancement factors.

Our shower forecast shares many features with an earlier analysis done by McBride \cite{1997AdSpR..20.1513M,2001indu.book..163M}. McBride based his shower list on \cite{1994Ana...287..990J}; our shower list was initially heavily based on the same paper and still retains many shower parameters from it. McBride converted ZHR to flux using the population index and a mass-magnitude-velocity relationship based on \cite{1967SCoA...10....1J}; we take a similar approach, albeit using the method of \cite{1990JIMO...18...44K} and \cite{1990JIMO...18..119K}. Like McBride, we also compare our results to the sporadic flux, although we do so for several kinetic energy thresholds rather than a single penetration threshold. We have added the computation of enhancement factors that express shower activity in terms of the factor by which the background activity is increased. Perhaps most importantly, our forecasts differ from that of McBride in that we update our shower list yearly to take into account both new measurements and model predictions of unusual activity.

\subsection{Shower activity parameters}
\label{sec:params}

There are no measurements of meteor shower fluxes from in-situ experiments. Thus, we must use observations of meteors in the Earth's atmosphere to quantify meteor shower activity. Meteor astronomers typically measure activity in terms of ZHR. ZHR is not simply a measure of hourly rate, but is rather the hourly rate of meteors that would be seen with the naked eye under ideal observing conditions by an alert observer when the shower radiant is directly overhead \cite{2008fgmm.book.....N}. All other observations -- those made during a full moon, say, or with a meteor radar rather than the naked eye -- must take their observing conditions and instrument sensitivity into account in order to calculate an equivalent ZHR. These ZHR values can then be used to compare showers or to compare a single shower's activity levels over time. If the cumulative number of meteoroids above some threshold mass follows a power-law, the shower ZHR is proportional to the shower meteoroid flux.

\begin{figure} \centering
\includegraphics{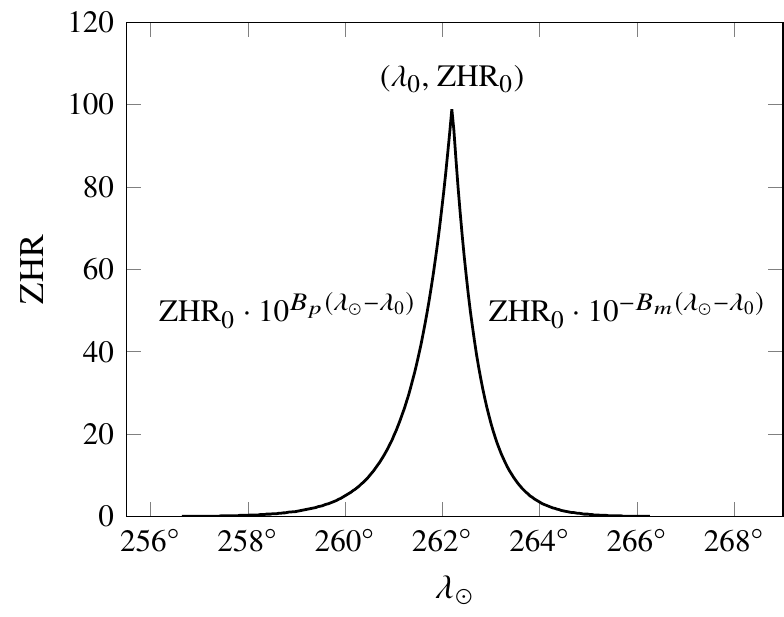}
\caption{The standard double-exponential shape of a meteor shower activity profile.}
\label{fig:diagram}
\end{figure}

Full ZHR profiles are needed because meteor showers do not have a well-defined duration.  Instead, a shower's activity tends to peak at a particular time each year and activity gradually builds up prior to the peak and dies down after the peak (see Fig.~\ref{fig:diagram}).  In most cases, this pattern of activity can be described by a double exponential function \cite{1994Ana...287..990J,1997AdSpR..20.1513M}.  Note that we express time in these activity profiles in terms of solar longitude, which simply measures the Earth's position in its orbit around the Sun. Peak activity for a particular meteor shower tends to occur each year at the same solar longitude, rather than the same day and time, due to the discrepancy between the length of a calendar year (either 365 or 366 days) and the Earth's orbital period (approximately 365.256 days). Different measures of shower duration -- such as a full-width at half-max -- can be constructed from these shower profiles, but in practice the ``duration'' of a potentially hazardous meteor shower depends on a spacecraft's sensitivity level as well as the shower's activity profile.

For the purposes of shower forecasting, we model the activity profile of each shower with four parameters: a peak zenithal hourly rate (ZHR$_0$), the solar longitude at which peak activity occurs ($\lambda_0$), and two exponents that characterize the shape of the shower's activity profile ($B_p$ and $B_m$).  Peak or maximum shower activity takes place when the Earth passes through the center or most densely populated region of the meteoroid stream. As illustrated in Figure \ref{fig:diagram}, ZHR increases with time (or rather, solar longitude) before the peak and decreases after the peak:
\begin{equation}
\mbox{ZHR} = \mbox{ZHR}_0 \cdot 
	\begin{cases}
	10^{+B_p(\lambda_\odot - \lambda_0)} & \lambda_\odot \le \lambda_0\\
	10^{-B_m(\lambda_\odot - \lambda_0)} & \lambda_\odot > \lambda_0
	\end{cases} \label{eq:shape}
\end{equation}
This double-exponential form is adequate for modeling the activity of most meteor showers. However, in some cases, such as the Perseids, we can better model the shower's activity by stacking two double-exponential profiles; in such cases we term the broader profile as the ``base'' and the narrower profile as the ``peak''.

\begin{figure*}
\includegraphics[width=\textwidth]{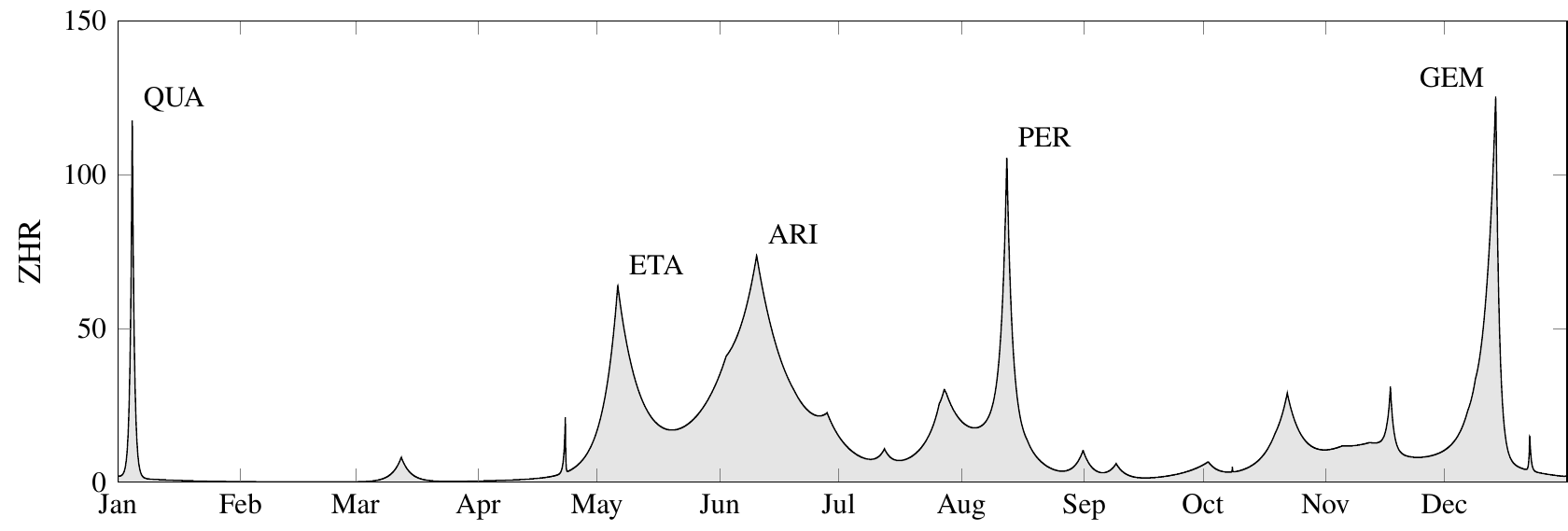}
\caption{ZHR from all active modeled showers over the course of an average year.}
\label{fig:zhr}
\end{figure*}

Our list of activity profile parameters is compiled from several sources. We rely in part on a set of ZHR activity profiles derived from naked-eye meteor observations \cite{1994Ana...287..990J}. We recently updated this list with parameters obtained by fitting double-exponential profiles to 14 years of meteor shower flux measurements \cite{MoorheadECSD} from the Canadian Meteor Orbit Radar \cite{2004ACP.....4..679W}. This paper represented the first significant improvement to the MEO's standard list of shower parameters in many years, and notable improvements were made for daytime meteor showers (which are visible only via radar) in particular. The full list of showers is presented in an Appendix; the total ZHR from these showers over the course of an idealized average year is presented in Figure \ref{fig:zhr}. The  following showers with a peak ZHR exceeding 50 are labeled: the Quadrantids (QUA), the eta Aquariids (ETA), the Daytime Arietids (ARI), the Perseids (PER), and the Geminids (GEM).

The MEO also routinely conducts numerical simulations  \cite{2004EM&P...95..141M,2008EM&P..102..285M} of meteoroid streams that either produce meteor showers with a level of activity that varies from year to year, produce occasional meteor outbursts or storms, or have the potential to produce new showers or storms. For instance, in preparation for the 2018 annual meteor shower forecast, we simulated the Perseids, Leonids, Ursids, Northern and Southern Taurids, Aurigids, Draconids, Andromedids, May Camelopardalids, June Bo\"{o}tids, and the ``Columbids'' (meteors potentially produced by comet C/2015 D4 (Borisov)). We also survey the literature for meteor shower predictions generated by other modelers. If the results suggest higher or lower activity than usual, we adjust our shower parameters to reflect this; we predicted increased activity in 2018 from the Ursids, Perseids, Leonids, eta Aquariids, Orionids, Draconids, and Andromedids \cite{forecast2018}. If the changes are significant, we sometimes choose to conduct additional studies, such as the Draconid work discussed later in this paper \cite{EgalLetter}.

Our shower parameters describe the activity level and timing of meteor showers as seen from Earth. However, spacecraft near the Moon or at other locations distant from the Earth may encounter the stream at a slightly different time. We therefore use the spacecraft's solar longitude to compute shower timing. A rule of thumb is that the apparent timing of a meteor shower can differ from that at Earth by up to a minute per 2000 km of altitude. Since the timing of a meteor shower is not known to the minute, it is unnecessary to take this effect into account for spacecraft in low-Earth orbit (LEO). For spacecraft in LEO or lunar orbit, we may substitute the solar longitude of the Earth or Moon, respectively, to compute timing. 

The parameters summarized in Fig.~\ref{fig:diagram} and Eq.~\ref{eq:shape} characterize the pattern of visual activity produced by a meteor shower at Earth. However, the brightness of a meteor is a function of both its speed and its mass. In order to compute fluxes for each shower to our desired limiting mass or kinetic energy, two additional parameters are required. The first is the shower's speed at the top of the Earth's atmosphere, which factors into the conversion from brightness (magnitude) to mass. The second is the shower's population index $r$, which describes the meteor brightness distribution for a given shower and allows us to scale the flux from one limiting mass to another. These parameters are included in our shower list in the Appendix, and Section \ref{sec:fluxcalc} describes how we use them to calculate fluxes.

\subsection{Shower fluxes}
\label{sec:fluxcalc}

As mentioned in the previous section, ZHR describes the rate at which a meteor shower produces visible meteors.  ZHR can be converted to meteor flux by taking into account observer biases and shower characteristics.  We use the methodology of \cite{1990JIMO...18..119K} to calculate the flux of meteoroids that have an absolute brightness (defined as their apparent brightness when viewed from a range of 100~km) of at least magnitude +6.5:
\begin{align}
f_{6.5} = \frac{
\mbox{ZHR}_0 \cdot (13.1 r - 16.5)(r-1.3)^{0.748}}{37200\mbox{ km}^2} \, ,
\label{eq:f65}
\end{align}
where ZHR$_0$ is the maximum ZHR and $r$ is the population index as described in Section \ref{sec:params}.
We omit the average perception factor, $k_{\mbox{\scriptsize avg}} \sim 1$, from our calculations. Because ZHR has units of hr$^{-1}$, this equation yields flux in units of km$^{-2}$ hr$^{-1}$.

The magnitude-limited flux produced by Eq.~\ref{eq:f65} can be converted to the flux of meteoroids that are one milligram in mass or larger as follows \cite{1990JIMO...18..119K}:
\begin{align}
f_{\mbox{\scriptsize mg}} = f_{6.5} \cdot r^{9.775 \log_{10}{
\left({
    \textrm{29 km s}^{-1}/v_\textrm{\tiny TOA}
}\right)
}} \label{eq:fmg}
\end{align}
This equation essentially makes use of Verniani's relationship \cite{1973JGR....78.8429V} between magnitude, mass, and velocity to calculate the meteoroid mass that produces a magnitude 6.5 meteor at the shower's speed.  The shower velocity is that at the top of the atmosphere (or rather, at the 100~km altitude at which meteors typically begin to ablate); shower meteoroids will have a different velocity in interplanetary space, before being accelerated by the Earth's gravity.

\begin{figure}\centering
\includegraphics{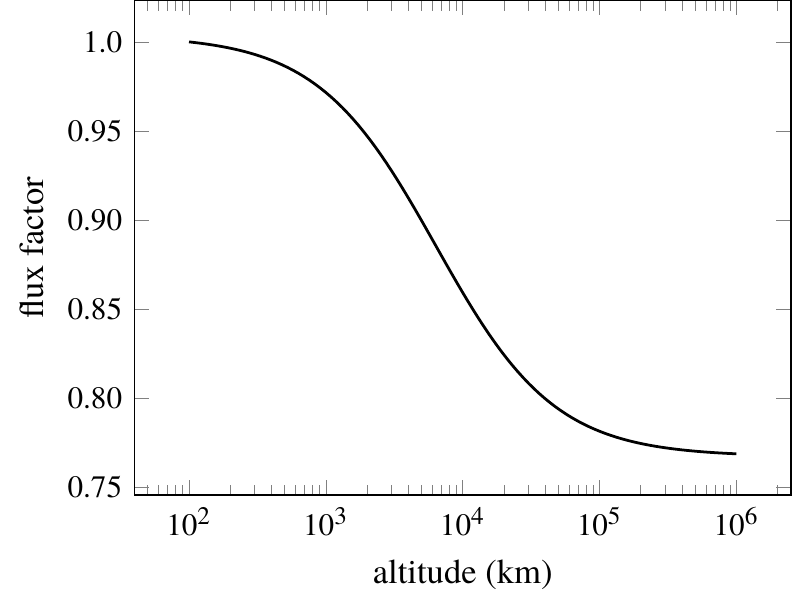}
\caption{Meteoroid flux, relative to that at the top of the atmosphere, as a function of altitude.}
\label{fig:gravDRA}
\end{figure}

The flux at the top of the Earth's atmosphere is not equivalent to the flux in interplanetary space. The Earth attracts and accelerates meteoroids, and so the meteoroid flux is enhanced near the Earth by what we term ``gravitational focusing.'' We must invert this gravitational focusing effect to obtain the flux at our desired altitude $h$ above either the Earth or the Moon \cite{Kessler:1972wn}:
\begin{align}
\frac{f_h}{f_\textsc{\tiny TOA}} &= \eta_g = \left({\frac{v_h}{v_\textsc{\tiny TOA}}}\right)^2 \, \mbox{, where}
\label{eq:grav} \\
v_h &= \sqrt{v_\textsc{\tiny TOA}^2 - \frac{2 G M_\oplus}{R_\oplus + \mbox{100 km}}
	+ \frac{2 G M_b}{R_b + h}} \, . \label{eq:vh}
\end{align}
In the above equations, $f_h$ denotes the flux of meteoroids at altitude $h$ and $f_\textsc{\tiny TOA} = f_{\mbox{\scriptsize mg}}$ denotes the flux of meteoroids at a height of 100 km above the Earth's surface. The variable $G$ refers to the gravitational constant, $M_\oplus$ the Earth's mass, and $R_\oplus = 6371$~km the Earth's average radius. The variables $M_b$ and $R_b$ refer to the mass and radius of the central massive body, which may be either the Earth or the Moon in our forecasts. Figure \ref{fig:gravDRA} gives this ratio as a function of altitude, assuming a meteoroid velocity of 23 km s$^{-1}$; the flux is greatest at the top of the Earth's atmosphere.  Note that Equation \ref{eq:grav} gives the average focusing factor. The exact focusing factor depends on the relative position of the shower radiant, massive body, and spacecraft; see \cite{2007MNRAS.375..925J} for a complete gravitational focusing algorithm.

\begin{figure} \centering
\includegraphics[width=0.4\textwidth]{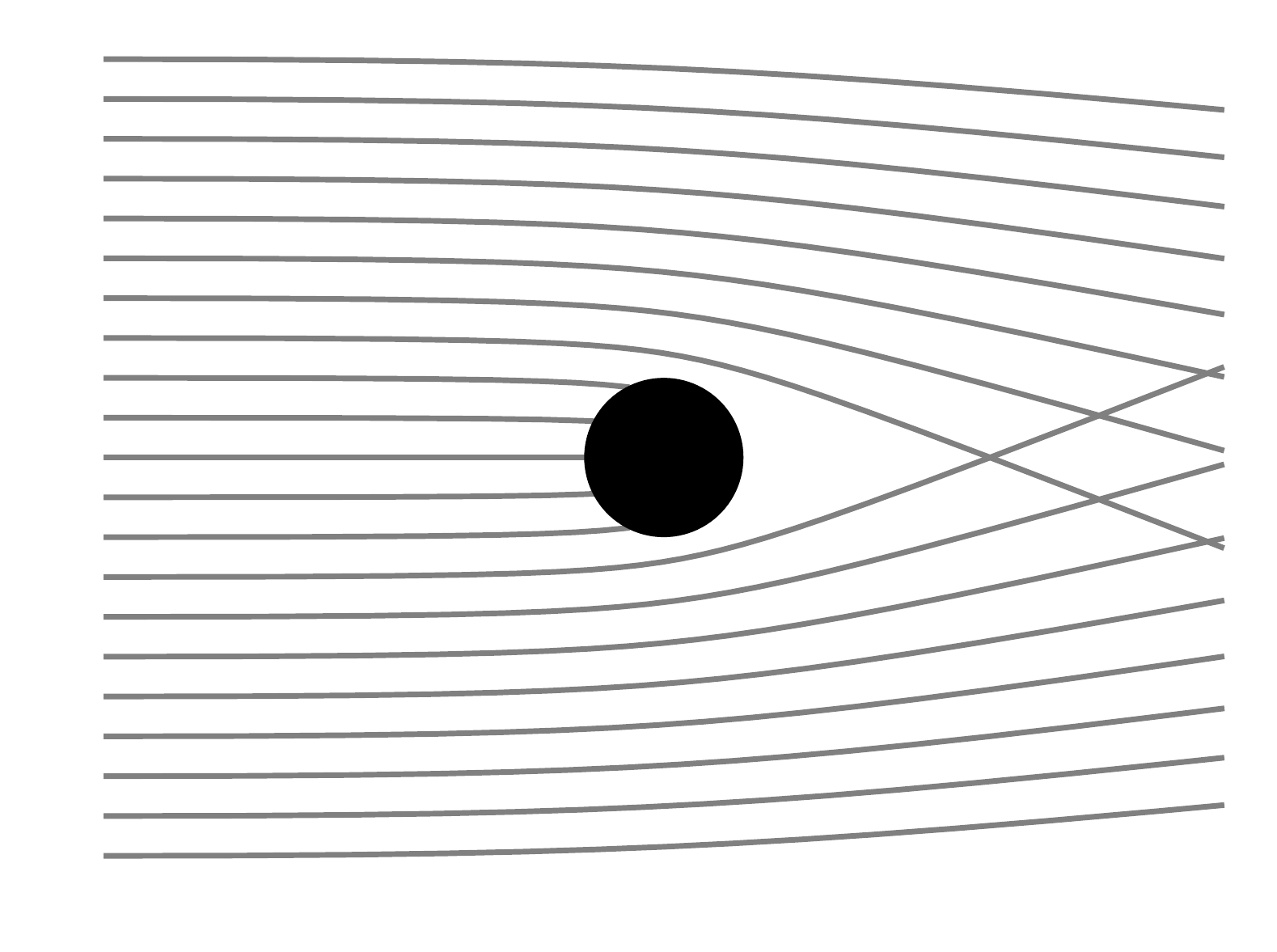}
\caption{An illustration of how meteoroid trajectories are bent due to the Earth's gravity.}
\label{fig:gravfocus}
\end{figure}

Spacecraft in low Earth orbit may also experience an effect called shielding, which occurs when the Earth or Moon physically blocks meteoroids from reaching the spacecraft (both gravitational focusing and shielding are illustrated in Figure \ref{fig:gravfocus}). For instance, if the Earth lies directly between the shower radiant and a low-orbiting spacecraft, the Earth can completely shield the spacecraft from the shower. We deliberately neglect this effect in our forecasts in order to report the flux corresponding to a ``worst-case scenario'' for shower exposure in which a spacecraft surface directly faces the shower radiant with no shielding. However, spacecraft operators can in some cases phase a satellite's orbit to use planetary shielding as a form of risk mitigation.

By incorporating gravitational focusing, we obtain the milligram-limited meteoroid flux at our desired altitude for a fully-exposed spacecraft surface. This flux can then be scaled to any arbitrary limiting mass using the relation:
\begin{equation}
\frac{f_{m}}{f_{\mbox{\scriptsize mg}}} = \left({\frac{m}{\mbox{1 mg}}}\right)^{1-s}
\end{equation}
where $s = 1 + 2.3 \log_{10} r$ is the shower mass index \cite{1990JIMO...18...44K}. 
We rarely report fluxes to a constant limiting mass, but instead usually report fluxes for a given limiting kinetic energy. The relationship between our kinetic energy threshold and our mass threshold is of course:
\begin{equation}
m_\textrm{lim} = 2 \, \mbox{KE}_\textrm{lim} / v_h^2 \, . \label{eq:duh}
\end{equation}
Section \ref{sec:limits} provides a full discussion of our chosen limiting quantities.

We typically combine the flux calculated for each meteor shower in our list to obtain the total flux due to all meteor showers as a function of time at a given location. In some cases, however, we have provided flux profiles on a shower-by-shower basis at a customer's request. We can also integrate fluxes over a time period to calculate shower fluence; this interval is typically six or seven hours.

\subsection{Limiting quantities}
\label{sec:limits}

Many ballistic limit equations (BLEs) are more closely related to kinetic energy than mass \cite{Hayashida:1991tm}.  Therefore, the primary limiting quantity to which we report fluxes is kinetic energy. When issuing forecasts for spacecraft in LEO, including the International Space Station (ISS), we report fluxes that correspond to the kinetic energy thresholds listed in Table \ref{tab:ms} \cite{Moorhead2017}.
Table \ref{tab:ms} also lists the mass and diameter that correspond to these kinetic energies when the speed relative to the spacecraft surface is 20 km s$^{-1}$ and the bulk density of the meteoroid is 1000 kg m$^{-3}$.

\begin{table} \centering
\caption{The four kinetic energies ({\normalfont KE$_\textrm{lim}$}) to which the MEO annual meteor shower forecast reports fluxes.  The second column lists the particle mass which, at 20 km s$^{-1}$, has the listed kinetic energy.  The third column lists the particle diameter which, for a bulk density of 1000 kg m$^{-3}$, has the listed mass.}\vspace{1em}
\begin{tabular}{r@{.}lcr@{.}l}
\multicolumn{2}{c}{KE$_\textrm{lim}$ (J)} & 
$m_\textrm{lim}$ at 20 km s$^{-1}$ & 
\multicolumn{2}{c}{$d_\textrm{lim}$ for 1000 kg m$^{-3}$} \\
\hline
   6&7~~ & $3.35\times 10^{-5}$ g & ~~~0&04 cm \\
 104&7 & $5.24\times 10^{-4}$ g & 0&1 cm  \\
 2827&  & $1.41\times 10^{-2}$ g & 0&3 cm  \\
104720& & $5.24\times 10^{-1}$ g & 1&0 cm
\end{tabular}
\label{tab:ms}
\end{table}

These values are chosen to be relevant to the International Space Station. A diameter of 0.04 cm is approximately the ballistic limit for spacesuit materials \cite{Cwalina2015}. Particle diameters of 1 mm and 3 mm are generally capable of penetrating delicate and robust spacecraft materials, respectively. A 1-cm particle will generally cause mission-critical damage\footnote{https://www.esa.int/Our\_Activities/Operations/Space\_Debris/\\Hypervelocity\_impacts\_and\_protecting\_spacecraft}.

Some of these thresholds are less relevant for unmanned spacecraft, such as the EVA suit puncture limit. When issuing forecasts that apply only to unmanned spacecraft (as we do in section \ref{sec:dra2018}), we may therefore substitute other limiting quantities. For instance, we present a size-limited flux for the 2018 Draconid forecast at the L1 and L2 Lagrange points (see Section \ref{sec:dra2018}). We select 100~$\mu$m, or 0.01 cm, as this limiting size, as particles of this diameter or slightly larger are capable of severing exposed wires. As an example, a 0.05~cm thick spring wire on the Hubble Space Telescope was cut by a particle impact, and it was asserted that particles 1/4 to 1/3 the wire thickness can cut wires \cite{2008AdSpR..41.1123D}.

\begin{figure*}
\includegraphics[width=\textwidth]{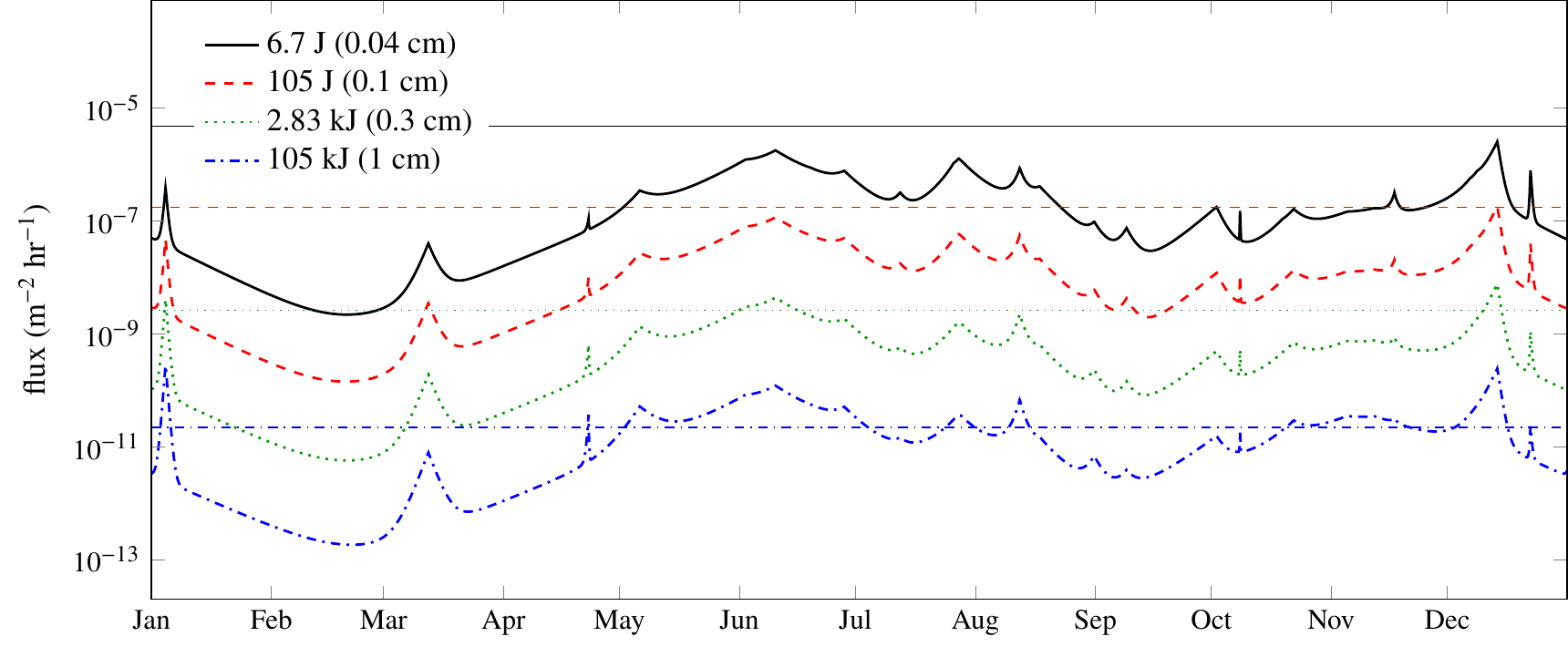}
\caption{Total shower flux (thick lines) and sporadic flux (thin lines) in an average year.}
\label{fig:flux}
\end{figure*}

Figure \ref{fig:flux} presents the total flux due to all modeled showers as a function of time within a ``typical'' year.  This example does not correspond to any particular year, but rather to an idealized year during which no shower shows unusual activity. This total flux corresponds to a spacecraft surface that directly faces each shower radiant (this is analogous to the ``perpendicular detector'' of \cite{1997AdSpR..20.1513M}, Figure 1). A real spacecraft surface cannot face multiple shower radiants at once, but one shower generally dominates at any given time, making this a useful estimate of activity. For each kinetic energy threshold, we also plot the Gr\"{u}n flux corresponding to that threshold. Note that in this idealized year, no shower exceeds the sporadic flux for our lowest energy threshold, but showers frequently exceed the sporadic flux at the highest energy threshold. This is because the particle size distribution is usually more shallow for showers than it is for the sporadic background; hence showers are proportionality ``richer'' in large particles. 

\subsection{Sporadic flux}
\label{sec:sporadic}

\begin{figure} \centering
\includegraphics{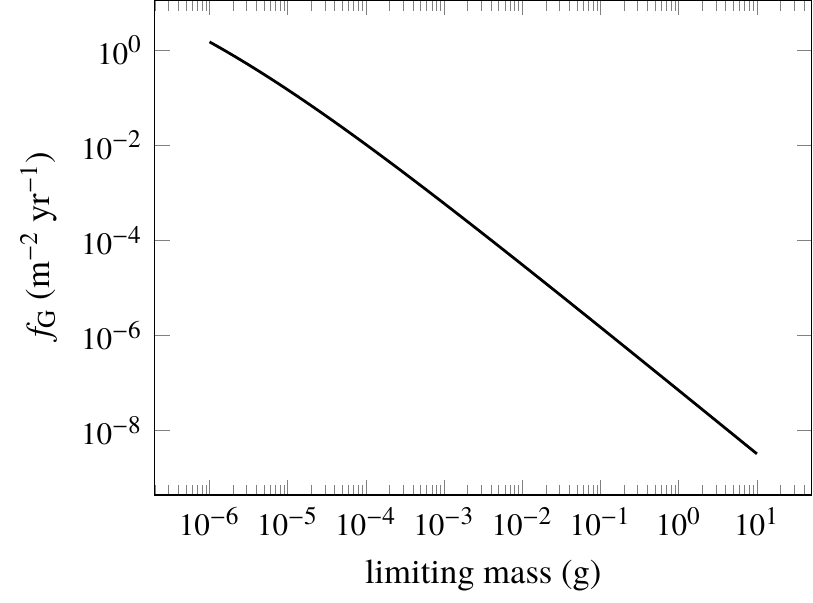}
\caption{The sporadic meteoroid flux on a randomly tumbling plate in interplanetary space at 1~au.}
\label{fig:fgrun}
\end{figure}

In order to facilitate risk assessments, the meteor shower forecast compares the total shower flux at a given time with the total time-averaged sporadic meteoroid flux.  We use the Gr\"{u}n et al.~model \cite{1985Icar...62..244G} for the interplanetary meteoroid flux near 1 au, which assumes a single sporadic meteoroid speed of 20 km s$^{-1}$. At large sizes, the Gr\"{u}n flux follows a power law of $f_\textrm{G} \propto m^{-1.34}$, but deviates from this power law at small sizes. This is apparent in Figure \ref{fig:fgrun} as a slight bend in the flux curve below $10^{-4}$ g. To obtain the sporadic flux at our desired altitude, we first calculate the sporadic speed at the given altitude using Equation \ref{eq:vh}. We use this speed to convert limiting kinetic energies to their equivalent mass at the given altitude. Alternatively, we assume a sporadic meteoroid density of 1~g~cm$^{-3}$ to convert limiting particle diameters to masses. Once these masses have been determined, we use Equation A3 of \cite{1985Icar...62..244G} to calculate the corresponding interplanetary sporadic flux.

The sporadic flux will also vary as a function of altitude due to gravitational focusing. However, as \cite{1985Icar...62..244G} provides the flux in interplanetary space, rather than at the top of the atmosphere, we use an alternate form of Equation \ref{eq:grav} that gives the gravitational focusing factor relative to the \emph{interplanetary} flux:
\begin{align}
\eta_g' &= 1 + \frac{2 G M_b}{R_b + h}\frac{1}{v_\textrm{ip}^2} \label{eq:grav2}
\end{align}
We assume an interplanetary sporadic speed of $v_\textrm{ip} = 20$~km~s$^{-1}$ for the purposes of calculating the gravitational focusing factor (Equation \ref{eq:grav}), which corresponds to $v_\textsc{\tiny TOA} = 22.87$~km~s$^{-1}$ at an altitude of 100~km above the Earth. 

In the case of sporadic meteors, we do assume that the spacecraft benefits from planetary shielding, in which the Earth or Moon blocks some meteoroids from reaching the spacecraft. On average, the shielding factor is \cite{Kessler:1972wn}:
\begin{align}
\eta_s &= \frac{1 + \cos{\psi}}{2} \label{eq:shield} \, \mbox{, where} \\
\sin \psi &= \frac{v_\textrm{TOB}}{v_h} \frac{R_b + H_b}{R_b + h} \, .
\end{align}
$H_b$ is the depth of the atmosphere that is capable of blocking meteoroids, which is 100~km for the Earth and 0~km for the Moon. Note that the shielding term $\eta_s$ also depends on meteoroid speed; some documents \cite{Smith:1994wa} cite a speed-independent approximation, but this is not strictly correct.  The speeds at the surface of the massive body and at arbitrary altitude $h$ are:
\begin{align}
v_\textrm{TOB} &= \sqrt{v_\textrm{ip}^2 + \frac{2 G M_b}{R_b + H_b}} \, \mbox{, and} \label{eq:vb} \\
v_h &= \sqrt{v_\textrm{ip}^2 + \frac{2 G M_b}{R_b + h}} \, , \label{eq:vh2}
\end{align}
where $H_b$ is the altitude at which the massive body's atmosphere begins to block meteoroids: 100~km for the Earth, and 0~km for the Moon. Note that we use the subscript TOB to denote that the velocity is taken at the surface of a massive body, including any atmosphere; when the massive body is Earth, $v_\textrm{TOB} = v_\textrm{TOA}$. 

\begin{figure} \centering
\includegraphics{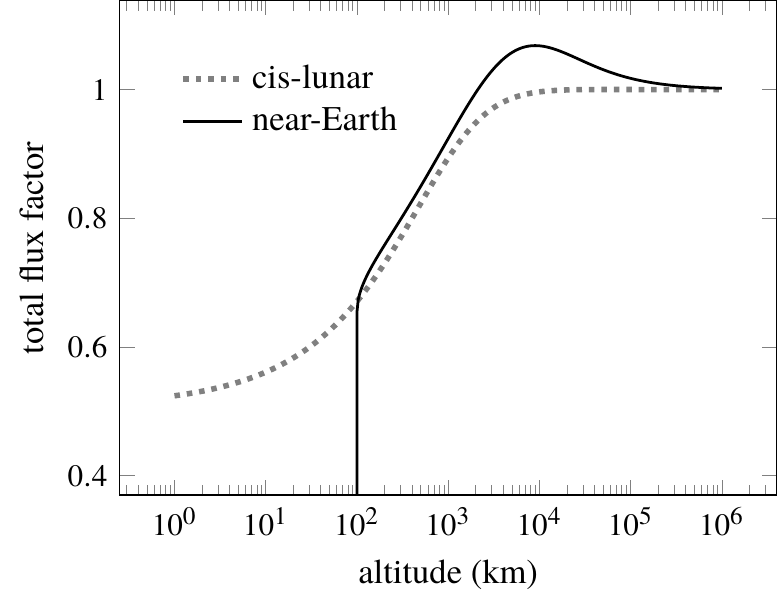}
\caption{Meteoroid flux, relative to that in interplanetary space, as a function of altitude (see Equation \ref{eq:totfac}).}
\label{fig:shield}
\end{figure}

Using Equations \ref{eq:grav2} and \ref{eq:shield}, we can convert the interplanetary flux we obtained using the Gr\"{u}n equation to a sporadic flux at our desired altitude:
\begin{align}
\frac{f_h}{f_\textrm{G}} &= \eta_g' \eta_s \label{eq:totfac}
\end{align}
Figure \ref{fig:shield} plots Equation \ref{eq:totfac} as a function of altitude above both the Earth (solid black line) and the Moon (dashed gray line). In the case of the Earth, gravitational focusing dominates at intermediate altitudes, increasing the overall flux, while shielding dominates at low altitudes. Shielding is the more relevant factor at all altitudes near the Moon due to the Moon's weaker gravity.

Our sporadic flux is a time- and position-averaged flux. We do not take into account the variations in gravitational focusing and shielding that apply to different radiant-planet-spacecraft configurations \cite{2007MNRAS.375..925J}. Neither do we take into account annual fluctuations in the sporadic flux \cite{2006MNRAS.367..709C,2009M&PS...44.1837C}. This total time-averaged flux was previously assumed to include both sporadic and shower meteoroids, prompting us to calculate and subtract the average shower component \cite{1997AdSpR..20.1513M,Moorhead2017}. 

We now believe that the Gr\"{u}n flux does not, in fact, contain contributions from meteor showers for the following reasons. The Gr\"{u}n cumulative flux curve in the meteoroid-size range (specifically those ranging from 100~$\mu$g to 100~g) was assumed to be proportional to $m^{-1.34}$ \cite{1985Icar...62..244G}, and Gr\"{u}n et al.~cite Whipple \cite{1967SAOSR.239....1W} for the exponent of -1.34. Whipple in turn states that the exponent of -1.34 is taken from a study of photographic meteors by Hawkins and Upton \cite{1958ApJ...128..727H}. Hawkins and Upton do appear to be the original source for the exponent of -1.34, and specify several times in their paper that only sporadic meteors were considered, and that shower meteors were excluded, in their analysis.

Thus, the \emph{slope} of the Gr\"{u}n flux curve in the meteoroid range was derived from a study of sporadic meteors.  
The \emph{amplitude} of the Gr\"{u}n flux curve is set by three in-situ experiments: Pioneer 8 and 9, HEOS, and Pegasus. Of these three, Pegasus detected the largest particles by far (the Pegasus experiment detected impacts corresponding to meteoroids approximately 10$^{-7}$ to 10$^{-6}$~g in size) and thus lies closest to our mass range of interest. However, a recent analysis of the Pegasus data revealed that the data show no signs of shower signatures (Blaauw, personal communication). Furthermore, most shower meteoroid streams are fairly eccentric, and small particles on eccentric orbits are likely to be preferentially removed from the Solar System by radiation pressure \cite{1979Icar...40....1B}. Because neither the slope nor the amplitude of the Gr\"{u}n flux curve incorporates shower meteoroids, we have now modified our forecasting algorithm to assume that the Gr\"{u}n flux describes sporadic meteoroids only.

\subsection{Enhancement factors}
\label{sec:enhance}

We do not typically directly report the sporadic fluxes described in Section \ref{sec:sporadic}. Instead, we report the ratio of the total or individual shower flux to the corresponding sporadic flux. These ``enhancement'' factors are intended to provide spacecraft operators with a quick assessment of how significant meteor shower activity is relative to the baseline sporadic activity. In previous forecasts, these enhancement factors could be negative when shower activity was below average \cite{MoorheadECSD}; now that we assume that the Gr\"{u}n flux consists only of sporadic meteoroids, our enhancement factors are uniformly positive.

\subsubsection{The average shower contribution fraction}

Gr\"{u}n's flux equation \cite{1985Icar...62..244G} reports the meteoroid flux incident on a randomly tumbling flat plate in interplanetary space.  The shower fluxes discussed in Section \ref{sec:fluxcalc} represent those on a plate facing the shower radiant; the zenithal hourly rate applies when the radiant is directly above the collecting surface.  This combination produces a worst-case scenario shower-to-sporadic flux ratio,\footnote{There is one conservatism that we do not include: gravitational focusing varies with spacecraft location, not just altitude, and can exceed the average factor we assume here.} in which the spacecraft surface faces the meteor shower and does not benefit from planetary shielding. In contrast, a surface facing away from the radiant, or located on the opposite side of the Earth or Moon from the radiant, would see no flux enhancement from the shower whatsoever. Thus, if one is to calculate the \emph{average} shower flux on a randomly oriented surface, one must include shielding and average over all possible orientations.

The intercepted shower flux is proportional to $\cos \theta$, where $\theta$ is the angle between a surface's normal vector and the shower radiant.  If we average $\cos \theta$ over all directions, and compare to the flux intercepted when $\theta = 0$, we obtain the following ratio:
\begin{align}
\frac{\oint \cos{\theta} \, \Pi(\tfrac{\theta}{\pi}) \, d \Omega}{\oint d \Omega}
&= \frac{1}{4} \, ,
\end{align}
where $\Pi$ is the Heaviside pi or rectangular function and ensures that particles cannot contribute by hitting the ``back'' of our one-sided plate.
Note that in this paper, we use $\oint d \Omega$ as a shorthand for 
$\int_0^{2\pi} \int_0^{\pi} \sin{\theta} \, d\theta \, d\phi$.

If we also include the reduction in flux produced by planetary shielding ($\eta_{\mbox{\scriptsize s}}$), the fraction of the total flux made up of shower meteors is:
\begin{align}
\alpha(h, \lambda_\odot) &= \frac{\tfrac{1}{4} \sum_i{f_{i,s}(h, \lambda_\odot)}}{f_\textrm{G}(h) + \tfrac{1}{4} \sum_i{f_i(h, \lambda_\odot)}} \, \mbox{, where} \label{eq:alpha} \\
f_{i,s}(h, \lambda_\odot) &= \eta_g(h, v_i) \, \eta_s(h, v_i) \, f_i(\mbox{100 km}, \lambda_\odot) \, \mbox{, and} \\
f_\textrm{G}(h) &= \eta_g'(h, v_\textrm{G}) \, \eta_s(h, v_\textrm{G}) \, f_\textrm{G}(h = \infty) \, .
\end{align}
Note that the shower fraction $\alpha$ is a function of altitude; because gravitational focusing and shielding are a function of both altitude and meteoroid speed, the relative strength of meteor showers with different speeds will vary with respect to each other and with respect to the Gr\"{u}n flux. The shower fluxes, and thus $\alpha$, also vary as a function of time or solar longitude.

\begin{figure} \centering
\includegraphics[]{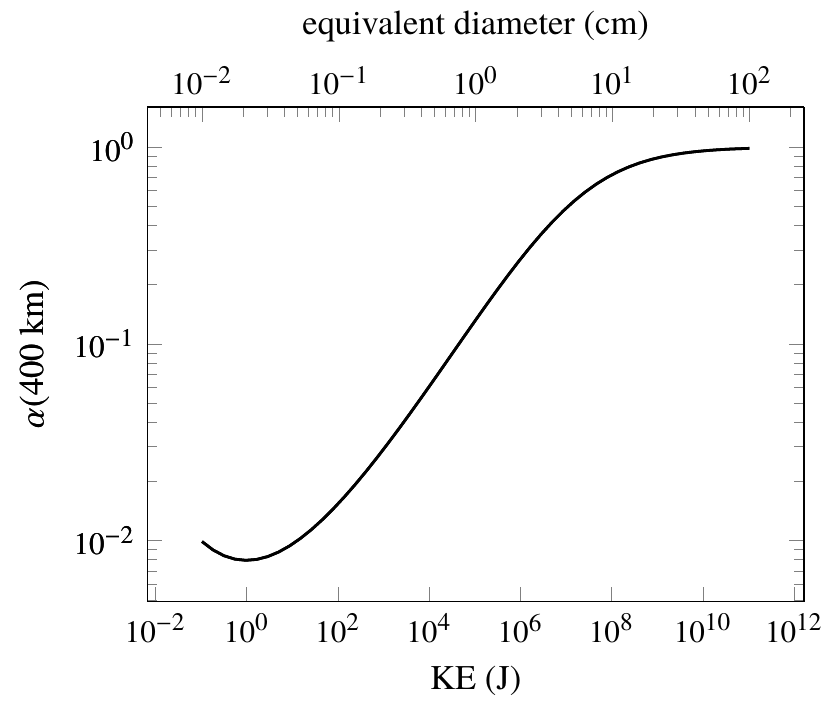}
\caption{Average shower contribution to the total meteoroid flux as a function of limiting kinetic energy.}
\label{fig:alpha}
\end{figure}

If we average Equation \ref{eq:alpha} over all 360$^\circ$ of solar longitude, we obtain the average shower contribution to the meteoroid flux. Figure \ref{fig:alpha} shows this time-averaged shower contribution at an altitude of 400~km as a function of limiting kinetic energy. The top axis displays the equivalent diameter, assuming a speed of 20~km~s$^{-1}$ and a density of 1~g~cm$^{-3}$. For low energies, the sporadic flux dominates and the shower fraction is quite small. Note that the shower fraction is inversely related to kinetic energy below 1~J; this is due to the turnover in the Gr\"{u}n sporadic flux at those sizes \protect\cite{1985Icar...62..244G}. At higher energies, the shower fraction is larger; the shower fraction is about 50\% for a limiting kinetic energy of 11.2~MJ, which corresponds to a particle diameter of 4.7~cm at a speed of 20~km~s$^{-1}$ and a density of 1~g~cm$^{-3}$. 

\subsubsection{Individual shower enhancement factors}

Individually, meteor showers may enhance the sporadic flux by up to a factor
\begin{align}
\beta_i(h,\lambda_\odot) &= \frac{f_i(h, \lambda_\odot)}{f_\textrm{G}(h)} \, \mbox{, where} \label{eq:betai} \\
f_i(h, \lambda_\odot) &= \eta_g(h, v_i) \, f_i(\mbox{100 km}, \lambda_\odot) \, \mbox{, and} \label{eq:betafi} \\
f_\textrm{G}(h) &= \eta_g'(h, v_\textrm{G}) \, \eta_s(h, v_\textrm{G}) \, f_\textrm{G}(h = \infty) \, .
\end{align}
Equation \ref{eq:betai} differs from Equation \ref{eq:alpha} in several ways. First, it represents an enhancement factor, not a contribution fraction, and thus shower fluxes do not appear in the denominator of Equation \ref{eq:betai}. We do not include shielding in computing shower fluxes for our worst case scenario, and therefore a shielding factor does not appear in Equation \ref{eq:betafi}. The spacecraft surface is assumed to face the shower radiant, and thus shower fluxes are not multiplied by a factor of $\tfrac{1}{4}$. The individual shower enhancement factors are summed to obtain the total shower enhancement factor at any given time:
\begin{align}
\beta(h,\lambda_\odot) &= \sum_i{\beta_i(h,\lambda_\odot)} \label{eq:btot}
\end{align}

\begin{figure*} \centering
\includegraphics[width=\textwidth]{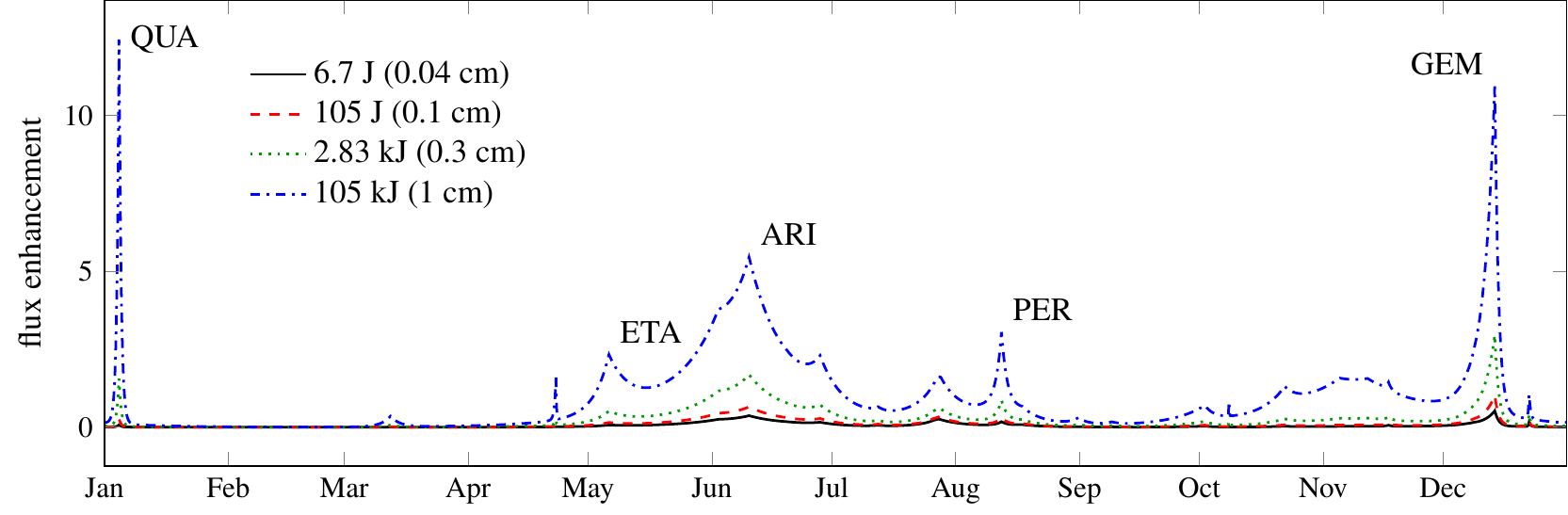}
\caption{The ratio of the total shower flux to the sporadic flux for four limiting kinetic energies.}
\label{fig:enhance}
\end{figure*}

\begin{figure} \centering
\includegraphics{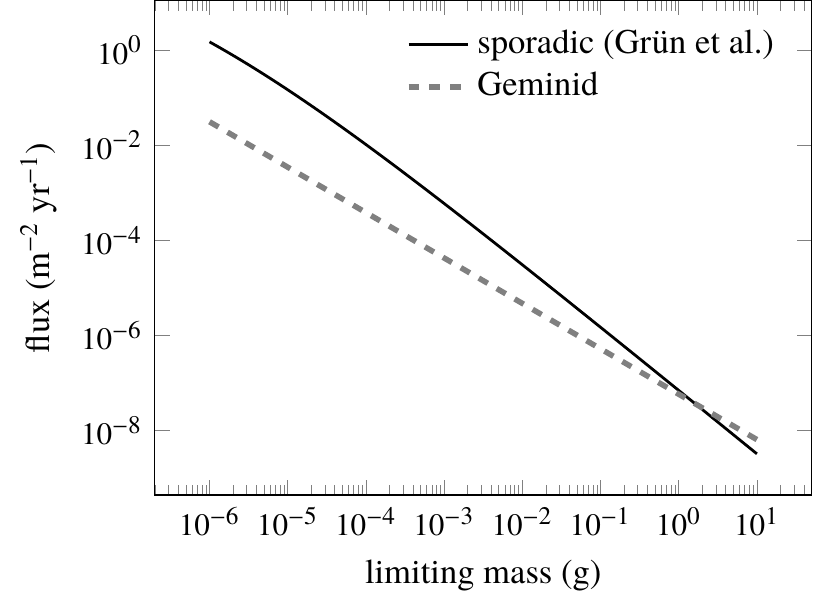}
\caption{Sporadic and Geminid meteoroid flux on a randomly tumbling plate at 1~au.}
\label{fig:comp}
\end{figure}

Figure \ref{fig:enhance} presents shower enhancement factors for a idealized year; we have assumed average activity for every shower, with no outbursts or storms. In such a year, the Quadrantids (QUA) and Geminids (GEM) produce the highest enhancements. Note that the enhancement factor tends to increase with particle size; this is due to the difference between the sporadic mass distribution and the shower mass distribution. The sporadic complex has a relatively ``steep'' mass distribution, meaning that the cumulative flux increases rapidly as we lower the limiting mass. In comparison, meteor showers have ``shallower'' mass distributions. An example is shown in Figure \ref{fig:comp}, in which we plot an estimate of the cumulative Geminid flux near peak activity (dashed gray curve) on top of the Gr\"{u}n sporadic flux (solid black curve). For small particles, the sporadic flux dominates, but for large particles, the Geminid flux eventually exceeds the sporadic flux. Thus, while sporadic and shower fluxes both increase as limiting mass decreases, shower fluxes increase less rapidly and thus the enhancement factor is smaller for less massive particles.

Our meteor shower forecasts report the total shower flux and total shower enhancement factor as a function of time, taking all active modeled showers into account.  These quantities are technically overestimates, as a flat surface cannot be normal to multiple radiants at once.  However, one shower usually dominates at any given time and thus our total flux and total enhancement factor provide useful estimates of the increase in meteoroid flux over the course of the year.  We do also provide radiant information for the strongest showers in a given year; if the enhancement factor or flux indicates that the risk of meteoroid impacts may exceed a missions's risk tolerance, the flux and enhancement factors can be used in combination with the radiant and the spacecraft's orientation to refine the shower risk assessment.

\subsubsection{Damage enhancement factors}
\label{sec:xtrafac}

Thus far we have presented size- and kinetic-energy-limited fluxes, fluences, and enhancement factors. Neither size nor kinetic energy is a function of impact angle; in contrast, both impact crater size and angular momentum transfer vary with impact angle. As a result, the shower enhancement factors for damage and attitude disturbance may exceed the enhancement factors corresponding to kinetic energy or linear momentum limits.

For instance, the modified Cour-Palais BLE for a single aluminum sheet takes the form \cite{Christiansen:1992cx}:
\begin{align}
p &= 5.24 d^{19/18} \textrm{BH}^{-1/4} (\rho/\rho_t)^{1/2} (v \cos{\theta} / c)^{2/3}
\label{eq:CPBLE}
\end{align}
where $p$ is the depth of the crater, $d$ is the diameter of the impact crater, BH is the Brinell hardness of the target material, $\rho$ is the density of the impactor, $\rho_t$ the density of the target material, $v$ the speed of the impactor relative to the target, $\theta$ the impact angle, and $c$ the speed of sound in the target material. A right-angle impact, according to Equation \ref{eq:CPBLE}, will penetrate deeper than an oblique impact. Thus, the ``worst-case scenario'' for impact risk enhancement due to a meteor shower occurs when the shower flux is perpendicular to a spacecraft surface, maximizing both the flux \emph{and} the crater depth. In this scenario, the shower is maximally damaging, especially in comparison with the sporadic flux, which will produce many oblique impacts that are less likely to penetrate the surface. 

In order to determine the ballistic limit enhancement factor, we invert Equation \ref{eq:CPBLE} and substitute $m = \tfrac{\pi}{6} d^3 \rho_m$ to obtain:
\begin{align}
m_{lim}
	&= \frac{\pi \rho}{6} \left[ \frac{p}{5.24} \textrm{BH}^{1/4} 
	\left( \frac{\rho_t}{\rho} \right)^{1/2} 
	\left( \frac{c}{v \cos{\theta} } \right)^{2/3} \right]^{54/19} \, .
\end{align}
This limiting mass can then be inserted into our flux equations, shower or sporadic, to obtain the corresponding mass-limited flux. 
If we assume $p = 0.1$~cm, BH$ = 90$, $\rho = 1.0$~g~cm$^{-3}$, $\rho_t = 2.7$~g~cm$^{-3}$, $c = 6.1$~km~s$^{-1}$, $v = 20$~km~s$^{-1}$, and $\theta = 0^\circ$, we obtain a limiting mass of 72~$\mu$g. We can also expand around $\theta = 0$ to obtain the limiting mass as a function of impact angle with all other parameters fixed:
\begin{align}
m_{lim} &= m_0 \cos^{-36/19}{\theta} \, .
\end{align}

\begin{figure} \centering
\includegraphics[]{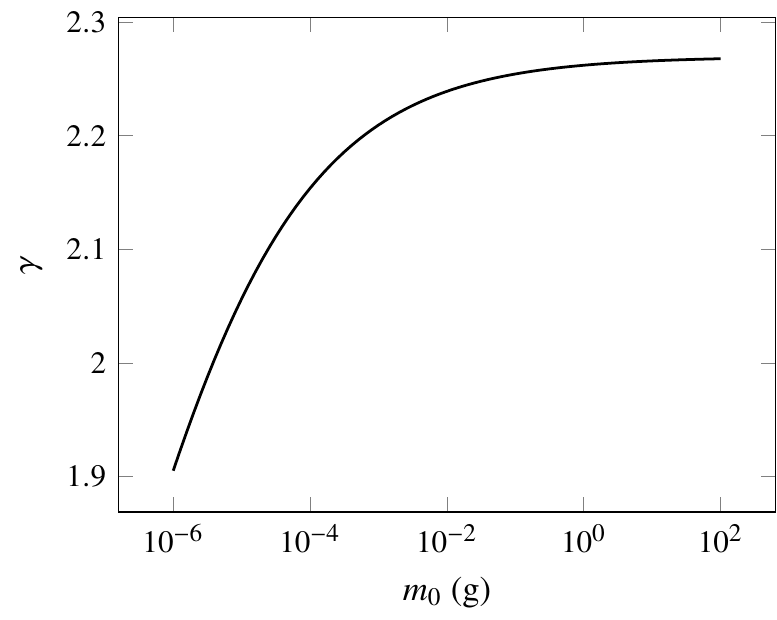}
\caption{Damage rate multiplier (see Equation \ref{eq:gamma}) as a function of limiting mass for $\theta = 0^\circ$.}
\label{fig:gamma}
\end{figure}

At any one point in time, shower meteoroids will all strike a flat spacecraft surface with the same angle of incidence.  However, our assumed isotropic sporadic flux will be distributed over all values of $\theta$, and thus $m_{lim}$ will vary.
To illustrate the significance of this effect, we consider the ratio of the mass-limited sporadic flux to the BLE-limited sporadic flux:
\begin{align}
\gamma &= 
\frac{\oint f_G (m_0) \, \Pi\left(\tfrac{\theta}{\pi}\right) \, \cos{\theta} \, d \Omega }{\oint f_G (m_0 \cos^{-36/19} \theta) \, \Pi\left(\tfrac{\theta}{\pi}\right) \, \cos{\theta} \, d \Omega }
\label{eq:gamma}
\end{align}
For our limiting mass of $m_{lim,0} = 72 \mu$g, we obtain $\gamma = 2.14$. The factor $\gamma$ is not a strong function of mass; within the 7 decades in mass that we typically consider the ``threat regime'' ($10^{-6}$ to 10~g), $\gamma$ is between 1.9 to 2.3 (see Figure \ref{fig:gamma}). Thus, we can obtain a crude estimate of the BLE-limited enhancement factor by multiplying our kinetic-energy-limited enhancement factors (see previous section) by 2.

\subsubsection{Attitude disturbance enhancement factor}
\label{sec:attitude}

Some missions must maintain a precise attitude; Gaia, for instance, requires an absolute pointing error of less than 1~arcmin \cite{2013Ana...551A..19R}. Meteoroids can therefore threaten a mission by introducing attitude disturbances. Angular momentum, like damage, has an additional dependence on impact angle that must be folded into considerations of the anomaly enhancement rate.

The angular momentum associated with a point mass is $L = m v \xi \cos \theta$; we can again invert this to obtain a limiting mass, which is 
\begin{align}
m_{lim} &= \frac{L}{v \xi \cos \theta} = m_1 \frac{\mbox{1 m}}{\xi \cos \theta} \label{eq:m1} \, ,
\end{align}
where $h$ is the angular momentum of the particle in the spacecraft's frame of reference and $\xi$ is the distance from the spacecraft's center of mass to the impact location.

\begin{figure} \centering
\includegraphics{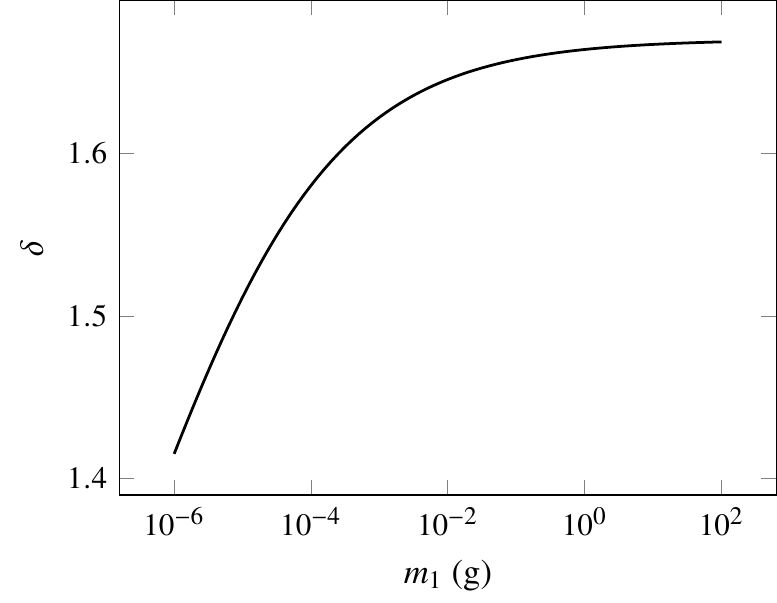}
\caption{Attitude disturbance rate multiplier (see Equation \ref{eq:delta}) as a function of limiting mass for $\theta = 0^\circ$.}
\label{fig:delta}
\end{figure}

An attitude disturbance rate multiplier can be derived by computing the ratio of the mass-limited sporadic flux to the momentum-limited sporadic flux:
\begin{align}
\delta &= 
\frac{\oint \int_0^{\xi_\textrm{max}} f_G (m_1) \, \cos{\theta} \, 2 \pi \xi \, d \xi \, d \Omega}{\oint \int_0^{\xi_\textrm{max}} f_G (m_1 \frac{\textrm{1 m}}{\xi \cos \theta}) \, \cos{\theta} \, 2 \pi \xi \, d \xi \, d \Omega} \label{eq:delta}
\end{align}
If we adopt $\xi_\textrm{max} = 5$~m to mimic Gaia's 10~m-diameter sun-shade, we obtain $\delta = 1.5$. Like the damage rate multiplier $\gamma$, $\delta$ is not a strong function of $m_1$ (see Figure \ref{fig:delta}). Thus, we can obtain a crude estimate of the attitude disturbance enhancement factor by multiplying the mass-limited enhancement factor by 1.5.

Note that for all individual shower enhancement factors, the impact angle is the angle between the surface normal and the meteoroid's direction of motion \emph{in the spacecraft's frame of reference.} The shower radiant itself is not adequate for calculating $\theta$; instead, one must calculate an aberrated radiant that includes the spacecraft's velocity vector. Note also that the momentum may be multiplied by a ``kickback'' effect. This momentum multiplier depends on the spacecraft material and is thought to vary from a factor of 1 to 3 \cite{Squire2017}; we have not included it in our analysis here.

\section{The 2018 Draconids}
\label{sec:dra2018}

The Draconids (also known as the October Draconids or Giacobinids) are a brief but occasionally intense meteor shower that occurs in early October. In most years, Draconid activity is very low or entirely absent. However, the shower has produced several storms (in the years 1933, 1946, and 2012), a number of outbursts (in 1926, 1952, 1985, 1998, 2005, and 2011), and reports of weak activity (in 1972 and 1996) \cite{2006Ana...451..339C, 2014mesh.book.....K}. Storm years produced zenithal hourly rates (ZHRs) of around 10,000; such rates correspond to a massive enhancement of the meteoroid flux. 

While several Draconid outbursts and storms have been correctly predicted -- including the recent 2011 outburst \cite{2013MNRAS.436..675Y} -- the Draconids have a history of being unpredictable.  For instance, activity in 1972 was much lower than anticipated and predictions of activity in the the 70s and 80s did not occur \cite{2014mesh.book.....K}. More recently, an unexpected outburst occurred in 2005 \cite{2006Ana...451..339C} and an unexpected storm in 2012 \cite{2014MNRAS.437.3812Y}.

The unpredictability of the Draconids may be the result of several factors. First, cometary activity is itself unpredictable and detailed historical light curves dating back hundreds of years simply do not exist. Furthermore, the orbit of the Draconid parent, comet 21P/Giacobini-Zinner, has been perturbed by Jupiter and by non-gravitational forces over time (most recently during the time period from 1966-1972; \cite{2014MNRAS.437.3812Y,1985cgha.book.....Y}) and we therefore lack a high-quality ephemeris for the comet prior to this period. Finally, the strength of the shower when it does occur can appear very different depending on the observation method. In both 2005 and 2012, the ZHR derived from radar observations was a factor of 4-5 higher than the reported visual rates \cite{2006Ana...451..339C,2014MNRAS.437.3812Y}. Because meteor radars detect smaller particles than most optical systems, this may indicate that the size distribution does not follow a simple power law. However, an alternative form for the size distribution has not been determined.

The possibility of enhanced Draconid activity in 2018 has been discussed by a number of modelers. Both Peterson \cite{PetersonBook} and Maslov\footnote{http://feraj.ru/Radiants/Predictions/1901-2100eng/Draconids1901-2100predeng.html} note that the geometry is favorable for enhanced activity in 2018, with the comet passing within 0.02~au of the Earth's orbit. However, geometry alone cannot be used to predict the level of activity; both Maslov and Jenniskens \cite{2006mspc.book.....J} note that the Earth may pass through a gap in the material, thus avoiding an outburst. Ye \cite{2014MNRAS.437.3812Y}, Kero and Kastinen \cite{2017P&SS..143...53K}, and Vaubaillon\footnote{http://www.imcce.fr/page.php?nav=en/ephemerides/phenomenes/ 
meteor/DATABASE/Draconids/2011/index.php} predict enhanced activity for 2018, albeit with some level of uncertainty.

\subsection{Simulation Results}

\begin{figure} \centering
\includegraphics[]{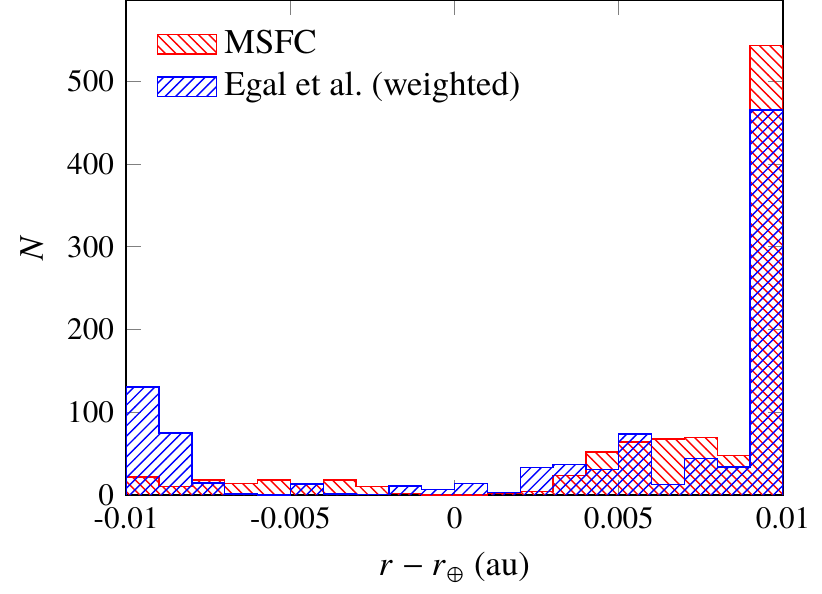}
\caption{Difference in heliocentric distance of Draconid meteoroid nodal crossings and that of the Earth.}
\label{fig:dists}
\end{figure}

Our forecast is based on two separate simulations of the Draconid meteoroid stream. We initially modeled the stream using the standard MSFC meteoroid stream model \cite{2008EM&P..102..285M}; additional simulations covering a wider range of distances were performed by Egal \cite{EgalLetter}. In agreement with Maslov and Jenniskens \cite{2006mspc.book.....J}, both sets of simulations predict that the Earth passes through a gap in the stream in 2018.  This is illustrated in Figure \ref{fig:dists}, which contains a histogram of the heliocentric distance between simulated Draconid meteoroid nodal crossings (the location where these meteoroids cross the ecliptic plane) and that of the Earth. These distances are positive when the particles cross outside the Earth’s orbit and negative when they cross inside. Results from both the MSFC meteoroid stream model \cite{2004EM&P...95..141M} and the Egal Draconid model \cite{EgalLetter} are depicted. Both models display a lack of particles within 0.002~au (300,000~km) of the Earth.

\begin{figure} \centering
\includegraphics[]{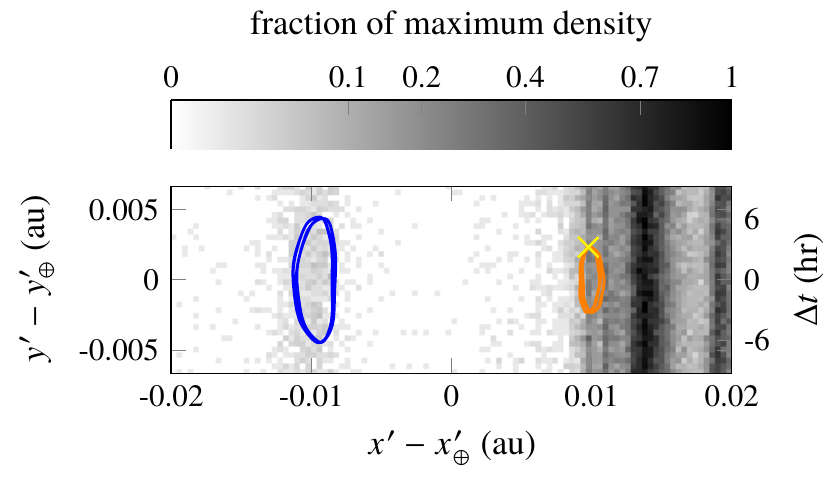}
\caption{SOHO and Gaia trajectories plotted overlaid on a heat map of Draconid nodal crossing locations.}
\label{fig:egal}
\end{figure}

Figure \ref{fig:egal} contains a heat map of simulated Draconid nodal crossings from \cite{EgalLetter} in the Earth’s corotating frame. Particles with a positive $y - y_\oplus$ value cross the ecliptic plane before the Earth reaches the stream and the inverse is true for particles with a negative value; we have plotted the equivalent time difference on the rightmost axis. The $x-x_\oplus$ coordinate is positive when the particle crosses the ecliptic plane at a larger distance from the Sun than the Earth. We have also included trajectories for the SOHO and Gaia spacecraft. SOHO's trajectory between January 1, 2015 and January 1, 2018 appears as a blue curve on the left half of the plot. Gaia's trajectory between June 20, 2016 and June 20, 2019 appears as an orange curve on the right half of the plot. Gaia's predicted location during the 2018 Draconid shower appears as a yellow ``x.'' SOHO's predicted location during the 2018 Draconid shower was unknown to the authors at the time of writing.

Visual meteor rates at the Earth were expected to be low; ZHR estimates ranged from 10-15 \cite{cal2018} to ``tens'' \cite{EgalLetter}. Observed visual rates in 2018 are estimated at ZHR$ \sim 100$, exceeding these estimates by a factor of a few.

While rates at Earth were predicted to be modest, all simulations indicate higher activity near the L1 and L2 Lagrange points. Meteoroids at L1 and L2 do not encounter the atmosphere and produce meteors and therefore the usual measure of ZHR does not strictly apply, but the level of activity is equivalent to a ZHR of about 5000, earning it the label of a meteor ``storm'' \cite{2006ChAna..30...61M,2014mesh.book.....K}. 

There are four active spacecraft orbiting near the Sun-Earth L1 point; these spacecraft are
the Solar and Heliospheric Observatory (SOHO), 
the Advanced Composition Explorer (ACE),
the Global Geospace Science (GGS) {\sl Wind} satellite, and
the Deep Space Climate Observatory (DSCOVR).
There is one spacecraft orbiting near the Sun-Earth L2 point: the Gaia space observatory. Gaia has reported large numbers of meteoroid strikes in the past,\footnote{https://www.newscientist.com/article/dn25925-galaxy-mappers-first-discovery-surprise-space-debris/} and thus might be the most susceptible to a Draconid outburst. We therefore focus on forecasting activity near the L1 and L2 Lagrange points. However, these spacecraft are not located precisely at the Lagrange points; instead, their halo orbits take them on wide orbits around L1 or L2, shifting the timing and activity level of the Draconids near the spacecraft (see Figure \ref{fig:egal}). Thus, we used Gaia's trajectory, rather than that of the L2 point itself, to extract meteoroid close approaches and forecast activity. Trajectories for SOHO, ACE, \emph{Wind}, and DSCOVR, however, were not available to the authors at the time of writing, and we therefore report activity for the L1 point itself. In situations like these, programs can contact the NASA Meteoroid Environment Office, provide a spacecraft ephemeris, and request a custom forecast.

\begin{figure} \centering
\includegraphics{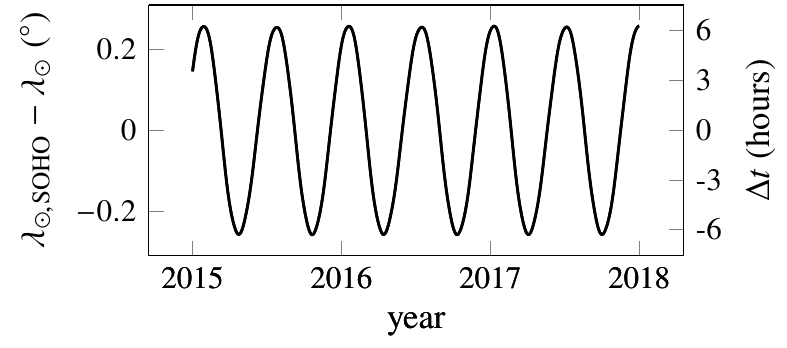}
\caption{Difference in solar longitude and hours between the SOHO spacecraft and the Earth.}
\label{fig:soho}
\end{figure}

We illustrate the potential shift in shower timing corresponding to changes in SOHO's position relative to the L1 Lagrange point in Figure~\ref{fig:soho}. This figures displays both the solar longitude offset between SOHO and the Earth in terms of both degrees (left axis) and hours (right axis). The spacecraft could encounter the Draconids up to 6 hours before or after the shower's peak time at the L1 point. We expect the potential shift in timing to be similar for other spacecraft orbiting L1. 

Draconid hourly rates are based on the Egal Draconid model \cite{EgalLetter}. We assume that Egal calibrated her predicted ZHR values to match historical data, and thus we make use of her ZHR values rather than her flux values. However, we do find that our top-of-atmosphere, magnitude-6.5-limited fluxes (see Eq.~\ref{eq:f65}) match those provided to us by Egal exactly. 

\begin{figure} \centering
\includegraphics{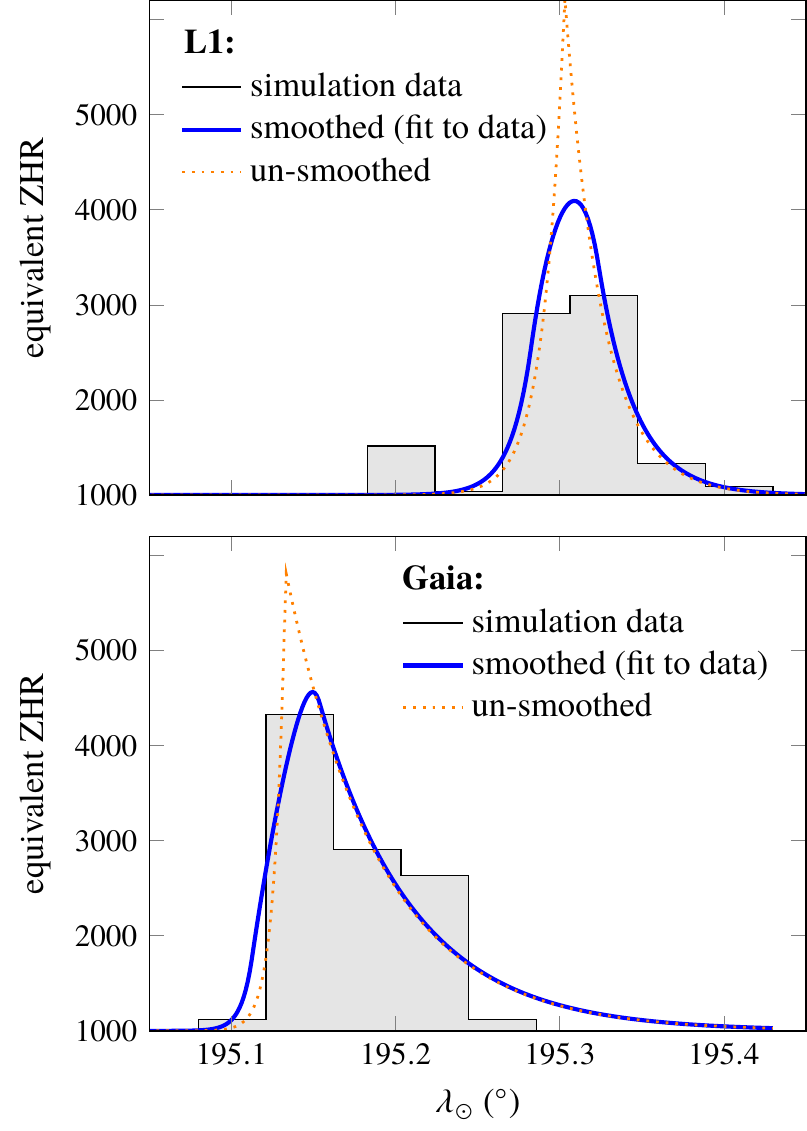}
\caption{Predicted 2018 Draconid ZHRs at the L1 Lagrange point and near Gaia.}
\label{fig:zhrL1}
\end{figure}

Egal et al.~modeled the Draconid activity by extracting particles that passed within an hour of the Earth. This approach was able to reproduce the sharp activity profile that is typical of the Draconid meteor shower, but reduces the number of contributing particles to 51 at L1 and 192 at Gaia near L2. In order to obtain the needed double exponential parameters, we have therefore fit not a double exponential, but a double exponential averaged over a 1 hour period. Figure~\ref{fig:zhrL1} displays predicted ZHR values for the 2018 Draconids as a function of solar longitude ($\lambda_\odot$) at the L1 Lagrange point and near Gaia. The Egal et al.~simulation results appear as a gray histogram. The blue curve represents our best fit of a smoothed double exponential function to the simulation results, while the dashed orange line represents the same double exponential function without smoothing. Explicitly including a 1-hour averaging period has a significant effect on the results. For instance, the peak ZHR near L1 appears lower than than near Gaia when considering the raw simulation results, but our fit indicates that the peak for L1 may simply appear lower due to bin placement.

Our forecast code then computes flux from ZHR following the methodology outlined in Section \ref{sec:method}. We use a population index of 2.6, a top-of-atmosphere velocity of 23~km~s$^{-1}$ (equivalent to 20.2~km~s$^{-1}$ at L1), and a density of 0.3~g~cm$^{-3}$ to scale the flux to the desired diameter or kinetic energy threshold. Finally, the flux is reduced to account for the much lower gravitational focusing at an altitude of 0.01~au (108,000~km).

Our choices of shower parameters are based on past measurements published in the literature \cite{Plavec1957,Kresak1975,1983Metic..18..386R,1986AJ.....91..159B,Simek1986,1994AnA...284..276S,1994CoSka..24..101G,PetersonBook,1995Ana...295..206J,1998JIMO...26..256A,Watanabe1999,Simek1999,2014EMnP..112....1M,2006Ana...451..339C,2006mspc.book.....J,2007AnA...473..661B,Koten2007,2008Icar..195..317B,Toth2012,Trigo2013,2013MNRAS.436..675Y,2014MNRAS.437.3812Y,Koten2014,Kac2015,2016Icar..266..331J,cal2018}, which are summarized in Table \ref{tab:draparms}. Generally, works cite a value for either the population index $r$ or the mass index $s$; similarly, a paper may quote a value for the meteoroid stream speed in interplanetary space $v_\textrm{ip}$ \emph{or} the speed at the top of the Earth's atmosphere, $v_\textrm{TOA}$. (We use slightly different nomenclature here than is typical in meteor astronomy, where $v_g$ is used in place of $v_\textrm{ip}$ and $v_\infty$ instead of $v_\textrm{TOA}$.) For the sake of comparison, we convert all values of $r$ to their equivalent values of $s$ and vice versa; we do the same for velocity.

\begin{table}
\begin{tabular}{cccccc}
$r$ & $s$ & $v_\textrm{ip}$ & $v_\textrm{TOA}$ & $\rho$ & Ref. \\
& & km~s$^{-1}$ & km~s$^{-1}$ & kg~m$^{-3}$ & \\
\hline
2.57 & \textcolor{gray}{1.94} & && & \cite{Plavec1957} \\
2.4-2.8 & \textcolor{gray}{1.9-2.0} & && & \cite{Kresak1975} \\
&& && 340 & \cite{1983Metic..18..386R} \\
&& && 200 & \cite{1986AJ.....91..159B} \\
& 1.99 & && & \cite{Simek1986} \\
\textcolor{gray}{2.89} & 2.06 &&&& \cite{1994AnA...284..276S} \\
\textcolor{gray}{3.04} & 2.11 &&&& \cite{1994AnA...284..276S} \\
&& 16.7 & \textcolor{gray}{20.0} && \protect\cite{1994CoSka..24..101G}\footnote{this paper cites a drastically different solar longitude (203.9$^\circ$, indicating that it may be referring to a different meteor shower} \\
&& \textcolor{gray}{20.1} & 23 & & \protect\cite{PetersonBook,1995Ana...295..206J} \\
3.0 & \textcolor{gray}{2.1} & && & \cite{1998JIMO...26..256A} \\
2.1 && && & \cite{Watanabe1999} \\
& 1.38 & && & \cite{Simek1999} \\
&& 17.4 && & \cite{2014EMnP..112....1M} \\
\textcolor{gray}{2.7} & 2.0 & 19.9 & \textcolor{gray}{22.8} && \protect\cite{2006Ana...451..339C} \\
&& 20.4 & \textcolor{gray}{23.2} && \protect\cite{2006mspc.book.....J} \\
&& & 23.57\footnote{based on one well-measured fireball} & 300 & \cite{2007AnA...473..661B} \\
& 1.78-1.87 & && & \cite{Koten2007} \\
&& 19.7 & \textcolor{gray}{22.6} && \protect\cite{2008Icar..195..317B} \\
2.62 && && & \cite{Toth2012} \\
2.3 && 20.76 && & \cite{Trigo2013} \\
\textcolor{gray}{2.12} & 1.75 & 17-19.1\footnote{
	Author describes this velocity as under-corrected for in-atmosphere deceleration
	} & - & 300\footnote{``consistent with''} & \protect\cite{2013MNRAS.436..675Y} \\
\textcolor{gray}{2.41} & 1.88 & \textcolor{gray}{20.5} & 23.27 & & \protect\cite{2014MNRAS.437.3812Y} \\
& 2.0 & && & \cite{Koten2014} \\
2.62 && & 22.4 & & \cite{Kac2015} \\
&& 20.7 & \textcolor{gray}{23.5} && \protect\cite{2016Icar..266..331J} \\
2.6 & \textcolor{gray}{2.0} & && & \cite{cal2018} \\
\end{tabular} \vspace{0.2in}
\caption{Draconid shower parameters (population index, $r$; mass index, $s$; geocentric speed in interplanetary space, $v_\textrm{ip}$; geocentric speed at the top of the atmosphere, $v_\textrm{TOA}$; and meteoroid bulk density, $\rho$) from the literature. Values in black are taken directly from the cited references, while values in gray are derived from the values in black.}
\label{tab:draparms}
\end{table}

\subsection{Forecast results at L1}

This section presents our forecast for the Draconid shower as encountered at the L1 Lagrange point. The basic properties of the shower are listed in Table \ref{tab:info}; note that the speed is less than it would be at the top of the atmosphere due to the lack of gravitational focusing from Earth at the L1 and L2 points. The shower peaks at 21:45 UT on October 8 and shows significant activity between approximately 20 UT and 24 UT (or 0 UT on Oct 9). All fluxes and enhancement factors apply to a spacecraft surface that directly faces the Draconid radiant. 

\begin{table}[b] \centering
\begin{tabular}{cc}
quantity & value \\
\hline
peak time at L1 & 2018-Oct-08, 21:45 UT\\
peak time at Gaia & 2018-Oct-08, 14:35 UT \\
velocity & 20.2 km s$^{-1}$ \\
radiant & R.A. 262$^\circ$, dec.~55$^\circ$\\
Earth-Sun-radiant angle & 82.6$^\circ$ \\
\end{tabular} \vspace{0.2in}
\caption{Properties of the Draconid meteor shower as seen from L1 and Gaia.}
\label{tab:info}
\end{table}

Table \ref{tab:flux} reports the maximum flux for four different particle thresholds. The first threshold is a limiting-size threshold of 0.01 cm; particles of this size may be capable of cutting wires that are between 0.03 and 0.04 cm in diameter \cite{2008AdSpR..41.1123D}. The remaining three thresholds are kinetic energy thresholds of 105 J, 2.83 kJ, and 105 kJ; these represent the approximate energy required to penetrate a delicate spacecraft surface, a typical spacecraft surface, and nearly any spacecraft surface, respectively. At a typical sporadic meteoroid velocity of 20 km s$^{-1}$ and a density of 1 g cm$^{-3}$ these energies correspond to particle diameters of 0.1, 0.3, and 1 cm. All thresholds and equivalent particle diameters are listed in Table \ref{tab:flux}.

\begin{table} \centering
\begin{tabular}{cccc}
particle & sporadic & Draconid & flux  \\
threshold & diameter & diameter & (m$^{-2}$ hr$^{-1}$) \\
\hline
0.01 cm & 0.01 cm & 0.01 cm & $2.4 \times 10^{-2}$\\
105 J & 0.1 cm &  0.15 cm & $1.1 \times 10^{-5}$\\
2.83 kJ & 0.3 cm & 0.45 cm & $4.7 \times 10^{-7}$ \\
105 kJ & 1 cm & 1.49 cm & $1.5 \times 10^{-8} $\\
\end{tabular} \vspace{0.2in}
\caption{The peak 2018 Draconid flux at the L1 point (rightmost column) for various particle thresholds (leftmost column). We report fluxes to one limiting particle diameter and three limiting particle energies; the second and third column of the table provide the equivalent particle diameter for sporadic and Draconid meteoroids, respectively.}
\label{tab:flux}
\end{table}

\begin{figure} \centering
\includegraphics[width=\columnwidth]{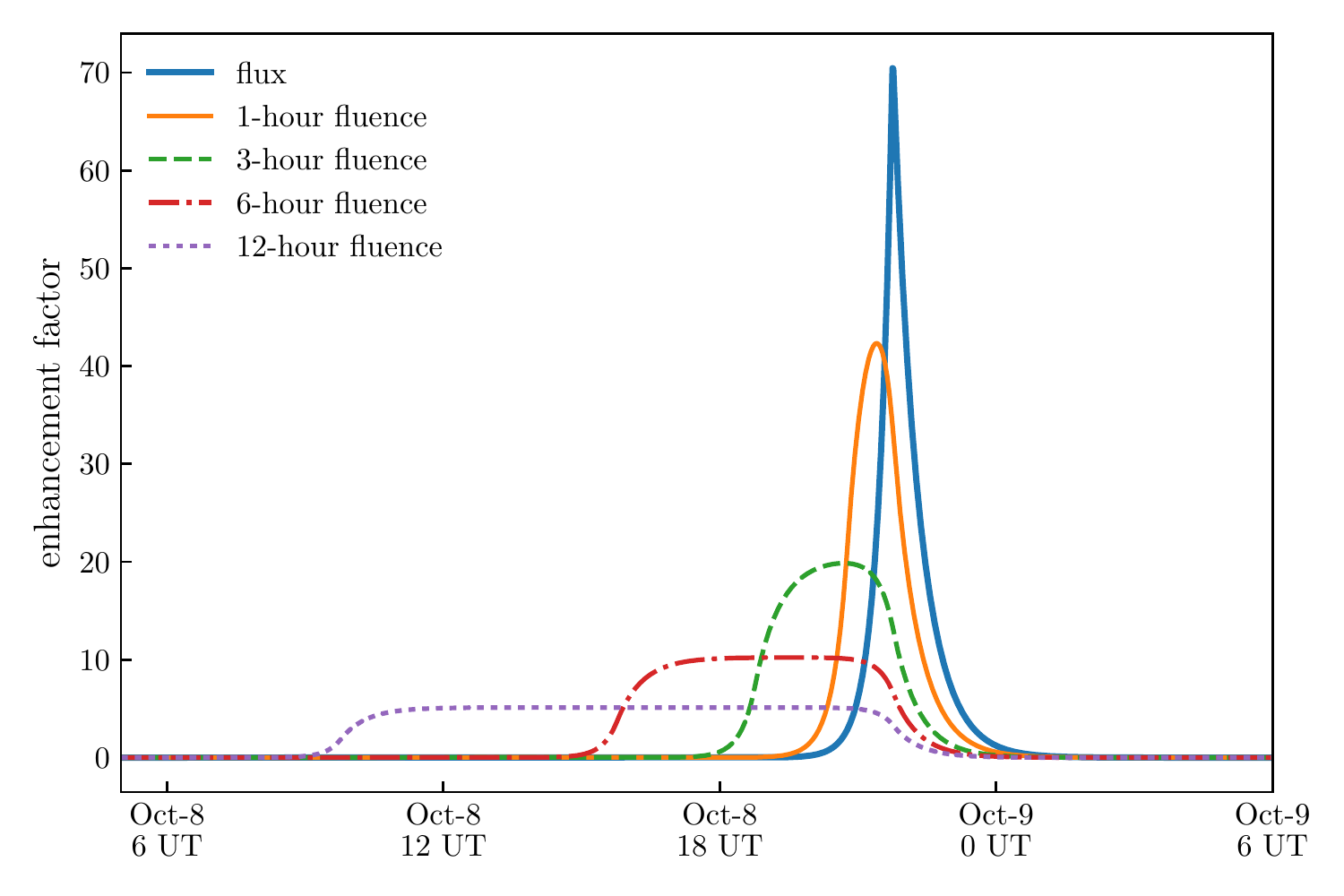}
\caption{The factor by which the sporadic flux or fluence is enhanced by the 2018 Draconids.}
\label{fig:factors}
\end{figure}

We have also calculated the fraction by which the typical sporadic flux is enhanced by the 2018 Draconids. Figure \ref{fig:factors} displays the flux enhancement factor for the 2018 Draconids as a function of time for our 105 J kinetic energy threshold. Because the shower is brief and sharply peaked, we also include enhancement factors for several different time intervals to illustrate how the significance of the shower varies with exposure interval. It's clear that short-term operations may benefit much more from mitigation measures than long-term operations.

\begin{figure} \centering
\includegraphics[width=\columnwidth]{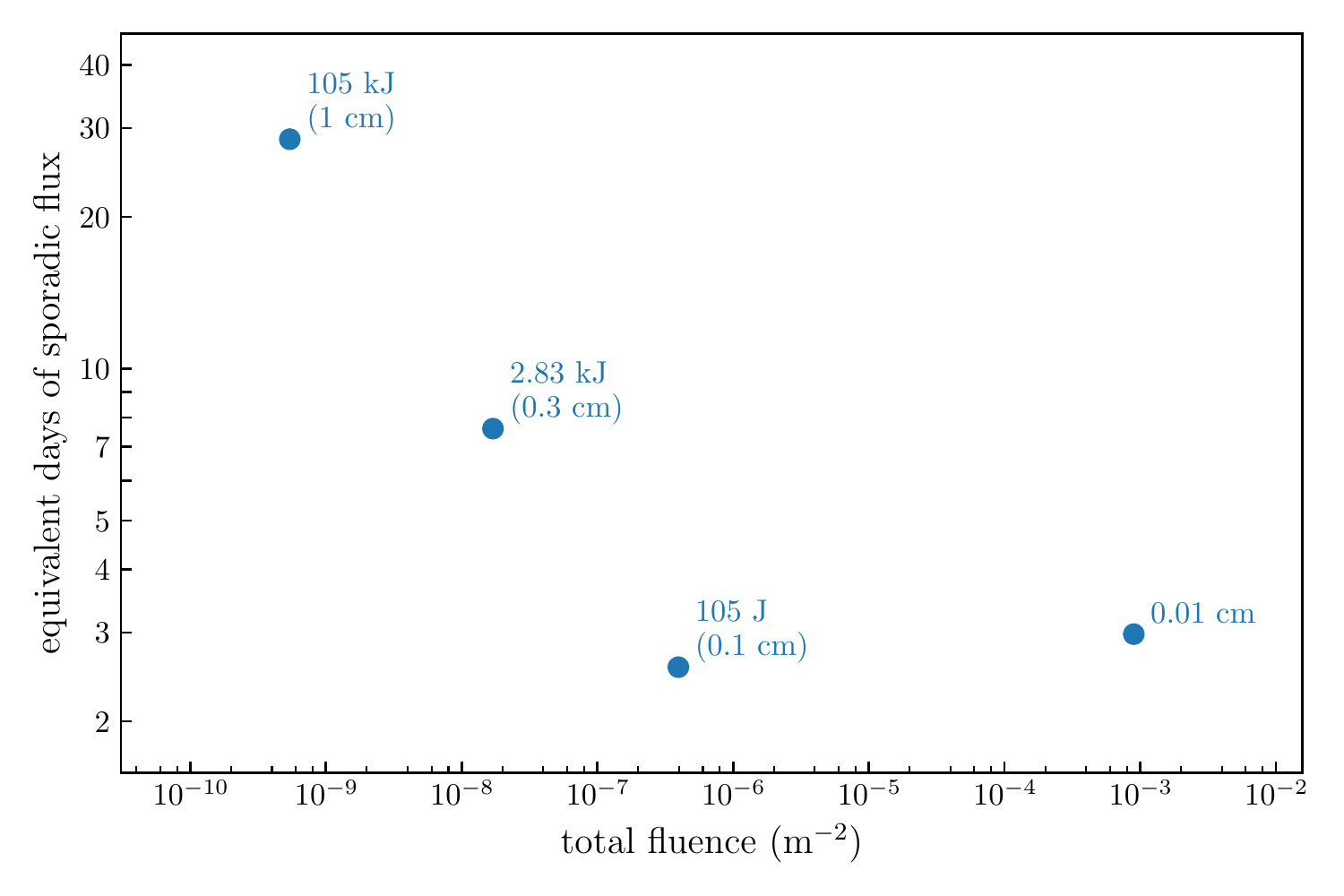}
\caption{The number of days of sporadic activity that equals the total 2018 Draconid fluence.}
\label{fig:fluence}
\end{figure}

Finally, we integrate over the entire duration of the shower to obtain the total number of Draconid particles encountered per square meter of area (Figure \ref{fig:fluence}). To provide context, we compute the time period during which such a surface would encounter an equivalent fluence from the sporadic meteoroid complex. This ranges from less than 3 days for the smallest energy threshold to about a month for the highest energy threshold. 

Both sporadic meteoroids and Draconids are more abundant at smaller sizes, although, like most showers, the Draconids are {\sl less} skewed toward small particles than the sporadic complex. Thus, while the Draconid flux is lower for larger particles, that lower flux corresponds to a larger ratio between Draconid and sporadic flux or fluence. 

Figure \ref{fig:fluence} presents the duration required for the sporadic fluence to equal that of the 2018 Draconids at the Sun-Earth L1 Lagrange point. These durations are presented for our four limiting quantities and plotted against the total fluence of Draconid particles. While the smallest threshold (0.01 cm) appears to break the inverse trend between fluence and enhancement factor seen in Figure \ref{fig:fluence}, this is simply the result of plotting energy-limited fluxes and a size-limited flux on the same graph.

\subsection{Forecast results at Gaia}

Gaia reportedy experiences frequent attitude disturbances due to meteoroid impacts, at a rate consistent with the Gr\"{u}n et al.~model \cite{1985Icar...62..244G,2016A&A...595A...1G}. 
Gaia's meteoroid-induced attitude disturbances are likely dominated by impacts on the spacecraft's large sunshield, which is 10.2~m in diameter. The threshold for detection is about $10^{-5}$~N~m~s; at a speed of 20~km~s$^{-1}$, an impact angle of 0$^\circ$, and a lever arm distance of 1~m, this corresponds to a limiting mass of $5\times 10^{-7}$~g. At the edge of the sun-shield, the limiting mass is just under $10^{-7}$~g.

\begin{figure} \centering
\includegraphics{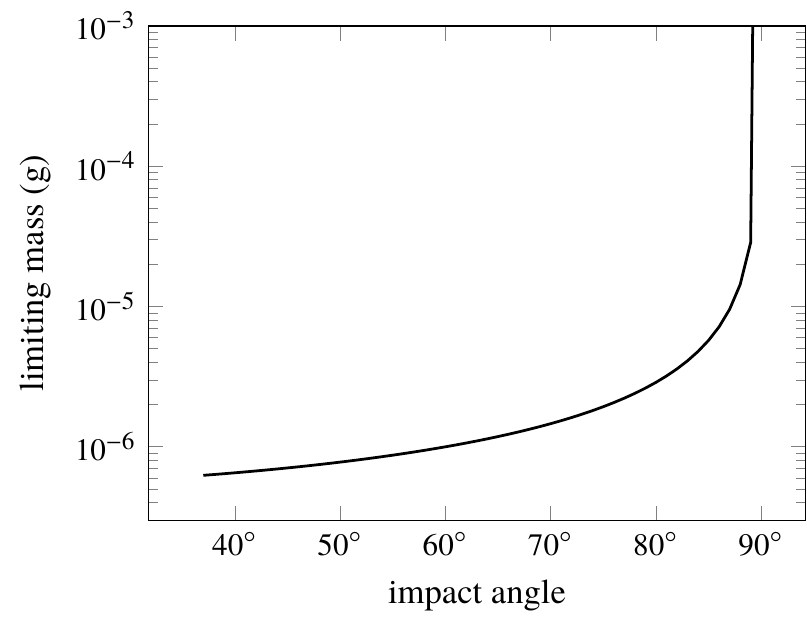}
\caption{The minimum Draconid mass required to produce a detectable attitude disturbance of the Gaia spacecraft.}
\label{fig:gaiamlim}
\end{figure}

However, the Gaia sun-shield does not face the Draconid radiant and will not experience the worst-case scenario that we typically model. Instead, the spacecraft maintains an angle of 45$^\circ$ between the sunshield and the Sun.\footnote{http://sci.esa.int/gaia/45313-deployable-sunshield/}  The aberrated (taking the spacecraft's velocity into account) radiant of the Draconid meteor shower has a Sun-centered ecliptic longitude of $\lambda - \lambda_\odot = 49.6^\circ$ and an ecliptic latitude of $\beta = 77.5^\circ$. Draconid impact angles range from 37$^\circ$ to 90$^\circ$ (angles exceeding 90$^\circ$ are treated as impacts of angle $180^\circ - \theta$ on the other side of the sunshield). This raises the limiting mass to at least $6.2 \times 10^{-7}$~g. For angles close to 90$^\circ$, a much larger impactor is required to produce a detectable attitude disturbance. Figure \ref{fig:gaiamlim} presents the limiting Draconid mass for a detectable Gaia attitude disturbance as a function of impact angle, assuming a detection limit of $10^{-5}$~N~m~s. The angular range shown corresponds to the range of possible impact angles of the shower on either side of Gaia's sunshield.

\begin{figure} \centering
\includegraphics{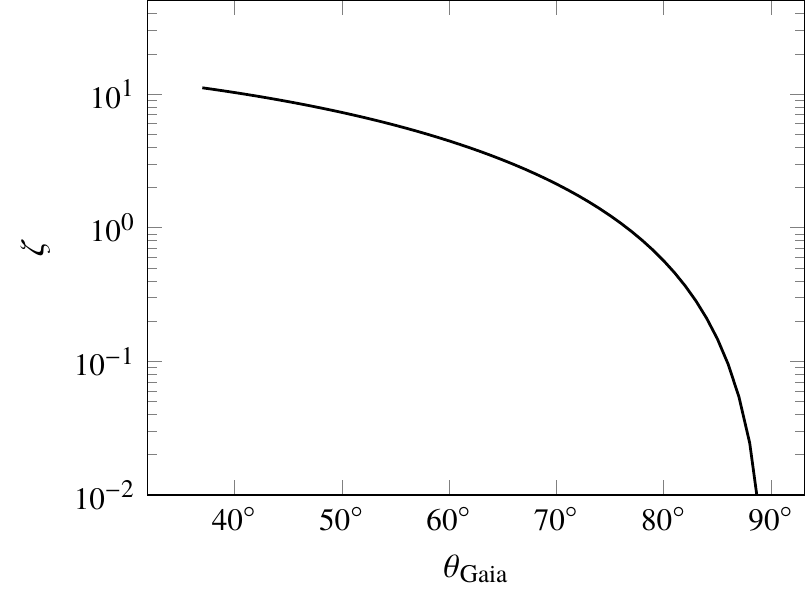}
\caption{Gaia attitude disturbance enhancement rate, $\zeta$, as a function of Draconid impact angle.}
\label{fig:gaiafactors}
\end{figure}

We analyze shower activity for Gaia as follows. First, we compute shower fluxes for Gaia using the angles and limiting masses shown in Figure \ref{fig:gaiamlim}. We apply a factor of $\cos \theta_\textrm{impact}$ to account for the fact that Gaia's sunshield will intercept fewer Draconids if the impact angle is high. We also calculate the sporadic flux for the limiting mass corresponding to a perpendicular impact, $5 \times 10^{-7}$~g. This can be converted to an attitude-disturbance-limited sporadic flux by dividing by $\delta = 1.5$; alternatively, the ratio of the shower to the sporadic flux can be multiplied by $\delta$:
\begin{align}
\zeta = \frac{f_{\textrm{imp},i}}{f_{\textrm{imp},G}} &= \delta \, \cos \theta_\textrm{Gaia} \, \frac{f_i(m_\textrm{lim}(\textrm{5.1 m}, \theta_\textrm{Gaia}))}{f_G(m_\textrm{lim}(\textrm{5.1 m}, 0^\circ))}
\label{eq:gaiafac}
\end{align}
This ratio is plotted in Figure \ref{fig:gaiafactors} as a function of the impact angle onto Gaia's sunshield, $\theta_\textrm{Gaia}$. If the impact angle is near its minimum, Gaia may see an order-of-magnitude increase in the number of attitude disturbances during the brief duration of the Draconid meteor shower. For an impact angle of about 76$^\circ$, the rate doubles (i.e., shower-associated disturbances equal sporadic-associated disturbances). For very large angles the enhancement is negligible.

\section{Conclusions and Future Work}

This paper describes a meteor shower forecasting model that can predict particle fluxes for spacecraft in low-Earth orbit as well as at other locations in near-Earth space, taking the gravitational pull and physical size of the Earth and Moon into account. We do not model secondary effects, such as lunar ejecta \cite{2015Natur.522..324H}, but can forecast primary impacts onto the lunar surface. We forecast fluxes assuming that the spacecraft surface in question directly faces an unobscured shower radiant; actual spacecraft surfaces may experience a reduced flux due to shielding or an oblique angle, or perhaps a further enhancement due to local variations in gravitational shielding.

We routinely compare our fluxes with the sporadic, or background, flux level; our flux enhancement factors may be used by spacecraft operators to quickly assess the possible increase in meteoroid impact risk associated with meteor showers as a function of time. We have developed the capability to generate kinetic-energy-limited, mass-limited, and diameter-limited fluxes and enhancement factors. These factors correspond to a worst-case-scenario for shower exposure in which the spacecraft surface directly faces an unobscured shower radiant. For this configuration, the damage or attitude disturbance rate may increase by a still larger factor; in general, we expect cratering rates to be increased by an additional factor of 2 and attitude disturbances to be increased by a factor of 1.5. All enhancement rates discussed in this paper will be lower if the spacecraft surface does not face the shower radiant.

We rely on both numerical simulations and historical observations to predict shower activity on the Earth and in low-Earth orbit. Simulations are particularly important for forecasts at very high geocentric altitudes, where the activity level of a particular shower may differ from that at the Earth. We present the example of the 2018 Draconids, which were predicted to be two orders of magnitude more active at the L1 and L2 Sun-Earth Lagrange points than at Earth itself. Spacecraft near the L1 point could experience up to a month's worth of typical meteoroid activity within a few hours, and the Gaia spacecraft could see a factor of 10 more attitude disturbances if the sunshield is tilted towards the shower radiant.

Forecasting cannot easily be extended outward from near-Earth space due to the lack of meteor shower data. Our shower parameters are in general derived from years, decades, or even centuries of meteor shower observations at Earth. However, the same set of meteoroid streams will not in general be encountered by spacecraft orbiting other planets. While a version of the Taurids, a stream of meteoroids originating from comet 2P/Encke that is also seen at Earth, is thought to impact Mercury \cite{2015GeoRL..42.7311C,2015Icar..250..230K}, comet C/2013 A1 (Siding Spring) produced a meteor shower on Mars \cite{2014Icar..231...13M,2014ApJ...792L..16K,2015GeoRL..42.4755S} but not at the Earth. One can build a list of possible streams and meteor showers for other planets based on comet orbits \cite{2004A&A...416..783S}, but translating close approaches into flux predictions is extremely difficult in the absence of any observations.

Meteor shower measurements near other planets are needed in order to generate shower lists equivalent to that in the Appendix; we are many decades away from collecting such data. For instance, while MAVEN (a Mars-orbiting NASA spacecraft studying the Martian atmosphere) detected atmospheric signatures from Siding Spring, typical meteor showers produce too little activity to be detected over the sporadic background. Thus, for the time-being, shower forecasts for other planets will require individual special efforts and carry enormous levels of uncertainty.

\section*{Appendix}
\label{apdx:list}

This appendix contains a list of those meteor showers of importance to the satellite impact hazard and the standard meteor shower activity parameters used as a starting point for annual meteor shower forecasts. Note that this list is a subset of the complete IAU Meteor Shower list which presently recognizes 112 established meteor showers.  For each shower, Table \ref{tab:list} provides a three-letter identification code; the solar longitude at which peak activity occurs, $\lambda_0$; peak zenithal hourly rate, ZHR$_0$; population index, $r$; activity growth exponent, $B_p$; activity decay exponent, $B_m$; and speed at the top of the Earth's atmosphere, $v_\textrm{TOA}$.

\begin{table*} \centering
\begin{tabular}{cccccccl}
ID & $\lambda_0$ & ZHR$_0$ & $r$ & $B_p$ & $B_m$ & $v_\textrm{TOA}$ & IAU shower name \\
& ($^\circ$) & (hr$^{-1}$) & & (deg$^{-1}$) & (deg$^{-1}$) & (km~s$^{-1}$) & \\
\hline
LYR & 33.15 & 18 & 2.1 & 1.229 & 4.620 & 49 
    & April Lyrids \\
ETA & 46.00 & 60 & 2.4 & 0.125 & 0.079 & 66 
    & $\eta$ Aquariids \\
ZPE & 72.36 & 20 & 2.7 & 0.028 & 0.028 & 29 
    & Daytime $\zeta$ Perseids \\
ARI &  79.72  &  60 & 2.7 & 0.065 & 0.055 & 39 
    & Daytime Arietids \\
SSG &  88.5   &   2 & 2.9 & 0.037 & 0.00 & 29 
    & Southern $\sigma$ Sagittariids \\
BTA &  96.7   &  10 & 2.7 & 0.100 & 0.00 & 29 
    & Daytime $\beta$ Taurids \\
PHE & 110.5   &   5 & 3.0 & 0.250 & 0.00 & 48 
    & July Phoenicids \\
PAU & 123.7   &   2 & 3.2 & 0.400 & 0.00 & 42 
    & Piscis Austrinids (peak) \\
PAU & 123.7   &   1 & 3.2 & 0.030 & 0.10 & 42 
    & Piscis Austrinids (base) \\
SDA & 124.93  &  20 & 3.2 & 0.109 & 0.071 & 42 
    & Southern $\delta$ Aquariids \\
CAP & 125.48  &   4 & 2.5 & 0.059 & 0.091 & 25 
    & $\alpha$ Capricornids \\
PER & 140.05  &  80 & 2.6 & 0.350 & 0.00 & 61 
    & Perseids (peak) \\
PER & 140.05  &  23 & 2.6 & 0.050 & 0.092 & 61 
    & Perseids (base) \\
KCG & 145.0   &   3 & 3.0 & 0.069 & 0.00 & 27 
    & $\kappa$ Cygnids \\
AUR & 158.6   &   9 & 2.6 & 0.190 & 0.00 & 66 
    & Aurigids \\
SPE & 166.7   &   5 & 2.9 & 0.193 & 0.00 & 64 
    & September $\epsilon$ Perseids \\
DSX & 189.44  &   5 & 2.7 & 0.063 & 0.167 & 32 
    & Daytime Sextantids \\
DRA & 195.438 &   2 & 2.6 & 7.225 & 0.00 & 23 
    & October Draconids \\
LMI & 206.14  &   2 & 2.7 & 0.093 & 0.079 & 61 
    & Leonids Minorids \\
ORI & 209.18  &  23 & 2.5 & 0.120 & 0.119 & 66 
    & Orionids \\
STA & 223.0   &   5 & 2.3 & 0.026 & 0.00 & 27 
    & Southern Taurids \\
NTA & 230.0   &   5 & 2.3 & 0.026 & 0.00 & 29 
    & Northern Taurids \\
LEO & 235.27  &  20 & 2.9 & 0.550 & 0.00 & 71 
    & Leonids (peak) \\
LEO & 234.4   &   4 & 2.9 & 0.025 & 0.15 & 71 
    & Leonids (base) \\
PUV & 255.0   &  10 & 2.9 & 0.034 & 0.00 & 40 
    & Puppid-Velid complex \\
MON & 257.0   &   3 & 3.0 & 0.250 & 0.00 & 42 
    & December Monocerotids \\
HYD & 260.0   &   2 & 3.0 & 0.100 & 0.00 & 58 
    & $\sigma$ Hydrids \\
GEM & 262.23  & 118 & 2.6 & 0.150 & 0.462 & 35 
    & Geminids \\
URS & 270.96  &  12 & 3.0 & 2.292 & 1.258 & 33 
    & Ursids \\
QUA & 283.58  & 120 & 2.1 & 0.865 & 0.827 & 41 
    & Quadrantids\\
GNO & 352.3   &   8 & 2.4 & 0.190 & 0.00 & 56
    & $\gamma$ Normids
\end{tabular}
\vspace{0.25in}
\caption{Activity parameters for the standard list of meteor showers included in a NASA Meteoroid Environment Office meteor shower forecast for near-Earth space.}
\label{tab:list}
\end{table*}

\section*{Funding Sources}

This work was supported in part by NASA Cooperative Agreement 80NSSC18M0046, by the Natural Sciences and Engineering Research Council of Canada, and by Jacobs contract S0MSFC18C0011.

\bibliography{local}

\end{document}